\documentclass[10pt]{article}
\usepackage{latexsym, amsmath, amsfonts, color, hyperref}
\usepackage{graphicx, float}
\newcommand{\noi}{\noindent}
\newcommand{\beq}{\begin{equation}}
\newcommand{\eeq}{\end{equation}}
\newcommand{\ba}{\begin{array}}
\newcommand{\ea}{\end{array}}
\newcommand{\bea}{\begin{eqnarray}}
\newcommand{\eea}{\end{eqnarray}}
\newcommand{\bi}{\begin{itemize}}
\newcommand{\ei}{\end{itemize}}

\newcommand{\ul}{\underline}
\newtheorem{rem}{Remark}
\newtheorem{prop}{Proposition}
\renewcommand{\baselinestretch}{1.3}

\begin{document}
\par \centerline{\bf \Large SDR Variance Estimates in Small Domains}
\medskip
\vspace{3mm}
\par\centerline{\bf by Eric Slud and Tim Trudell} \medskip
\par\centerline{University of Maryland, and US Census Bureau\footnote{This report is released to inform interested parties of ongoing research and to encourage discussion of work in progress. Any views expressed on statistical, methodological, 
technical, or operational issues are those of the authors and not those of the U.S.~Census Bureau.}}
\vspace{3mm}
  
\noi {\bf Abstract:} Successive Difference Replication (SDR) is a replication-based method of variance estimation 
introduced by Fay and Train (1995) for estimators based on complex multi-stage surveys, especially those 
including a final systematic-sampling stage. The method has been used for many years as the primary 
variance-estimation methodology in large national household surveys administered by the Census Bureau, 
including the American Community Survey and also the Current Population Survey's monthly estimates based 
on self-representing strata. In settings where it is applied, generally no second method of variance estimation 
has been available, so the performance of SDR has been studied via simulation by various authors, for variances 
of survey totals and of nonlinear survey estimators. Yet previous simulations have largely been based on estimation 
from simple random samples drawn from complete populations with attributes generated independently and 
identically distributed ({\tt iid}) from some distribution. The performance of the method in domain estimation based on 
non-SRS samples, for example in estimation of survey-weighted totals for small domains, has not been 
systematically studied, despite the common use of SDR-based domain variances in Small Area Estimation 
applications. This paper begins with a thorough exposition of the SDR method and review of previously published 
results on the small-domain biases of SDR variance estimation. It is shown that the number of $D$ cycles used 
in implementing SDR should be 3 or larger, in order to control the variability of SDR estimates, but need not be 
larger than 5; beyond that, the value of $D$ is virtually irrelevant to the occurrence of small-domain bias in SDR. 
The SDR method is shown via theoretical formulas and simulation to inflate average estimated variances in 
small domains by amounts that vary systematically with the patterns of attribute means and variances and survey 
weights in consecutively enumerated strata. The degree of average variance inflation is generally moderate, no 
more than 15\% in domains with sample size 20, but can be larger in special settings. Moreover, SDR estimates are extremely variable in small domains, with standard deviations often far larger than any biases.

\noi {\it Key words:\/} American Community Survey, Balanced Repeated Replication, Linear Algebra, 
Simulation Study, Small Area Estimation  

\section{Introduction} \label{Intro}

Successive Difference Replication (SDR) is a replication-based method of variance 
estimation introduced by Fay and Train (1995) for estimators based on complex multi-stage surveys, especially 
those including a final systematic-sampling stage. The method was specifically designed for applicability to the 
Current Population Survey (CPS) self-representing (SR) strata to replace a previous method by which SR PSU's 
had been split into half-samples for use with Balanced Repeated Replication (BRR) variance estimation. SDR 
has been used for many years as the primary variance-estimation methodology in large national household 
surveys administered by the Census Bureau including the largest one, the American Community Survey (ACS), 
and also the CPS monthly estimates based on SR strata. In this Introduction, we present the basic definition of 
the method, including also the multi-cycle variants of the method suggested by Fay and Train (1995) and 
implemented at the Census Bureau (Navarro 2001).

\subsection{Background and Basic Definitions} \label{basics}

Fay (1984) introduced into survey methodology a linear-algebra viewpoint on variance estimation by replication 
methods using quadratic-form variance estimators (not necessarily design-unbiased) in terms of weighted 
attributes from complex surveys. This approach was meant for application to 
large, national surveys with various designs often including a final-unit systematic-sampling stage. For reasons 
of logistical convenience and simplicity, national multistage household surveys often sample from geographic 
Primary Sampling Units using a systematic sampling design applied to a geographic within-PSU address list. 
For such sampling plans, there is no unbiased purely design-based variance formula even when response is 
complete, which it never is. Therefore, survey methodologists developed several variance estimation formulas, 
quadratic forms in the vector of sample-respondent weighted attribute of interest, that were only slightly 
biased. These variance estimates, so-called successive-difference estimates among them, were investigated 
by Wolter (1984) and assembled into a single useful source in the first edition of Wolter's (2007) book. 
Fay's (1984) approach, later augmented to Fay and Train's 1995 Successive Difference Replication,  enabled 
replication-based variance estimation which often 
was nearly unbiased and required only final (response-adjusted) weights, not joint inclusion probabilities 
which would be very hard to come by, for sample respondents to complex surveys.

The Fay methodology served as theoretical underpinning for the custom VPLX software\footnote{This software 
and associated documentation, now mainly of historical interest, are preserved at the Census Bureau website 
{\tt https://www.census.gov/data/software/vplx.html}.} that he coded for the US Census Bureau (now superseded 
by SAS, R and Python code written for specific surveys at the Census Bureau and by the WESTAT {\tt WesVar} 
package). One of the undocumented aspects of this replication software was the general quality of 
approximation of quadratic-form variance estimates by replicate-based variance estimates based on fixed 
numbers (generally 80 or 160) of replicates when those numbers were much less than the respondent sample 
sizes, although there were many later simulation studies verifying that these approximations were generally 
very good (accurate for variances of totals to within at most a couple of percent) on large surveys.

A complex probability sample survey results in a respondent sample of attributes $\{y_i\}_{i \in {\cal S}}$ 
observed for individuals with known (final, nonresponse-adjusted) weights $w_i$, from which it is desired to estimate 
the population total $t_Y$ of $y$-attributes with the estimator $\hat{Y} = \sum_{i\in {\cal S}} \, \breve{y}_i$, 
where from now on we use the symbol $~\breve{y}_i \, \equiv \, y_i \, w_i~$ to denote the survey-weighted 
attribute. We consider replication-based methods of estimating the design variance of $\hat{Y}$, that is, the 
variance of the random variable $\hat{Y}$ that takes all possible values based on the set of possible samples 
drawn, according to their respective probabilities, together with the unknown pseudo-randomization
 model according to which sampled individuals independently make the binary choice whether to respond or 
not. {\it No previous studies have in any way addressed the behavior of the SDR variance estimator due to 
survey nonresponse, and this paper also ignores the mechanism of nonresponse, adopting the usual 
convention of treating the final nonresponse-adjusted weights $w_i$ as the reciprocals of first-order 
inclusion probabilities $\pi_i$ with which individuals $i$ indexed are sampled and respond to the survey.\/}

Suppose that the frame population has been ordered according to a variable (usually geographic) known for 
all population members. Indices $i$ refer to individuals indexed within the frame 
population, while ordered (respondent) units in samples of fixed size $n$, denoted $i(j)$ for $j=1,\ldots,n$, are 
indexed by $j$. Throughout the paper, we maintain the \vspace{3mm} \\ 
\hspace*{5mm} {\bf Notation:} ~for any population quantity \ $(b_i, \; i \in {\cal U})$, ~the sample-indexed 
quantity is   \vspace{-2mm}
\beq b_j^{\circ} \; \, \equiv \; \, b_{i(j)} \qquad \mbox{for} \qquad j=1,\ldots, n  \label{not.circ} \eeq
This notation provides a distinction important for correctness of assertions about variance estimation.

Fay and Train (1995) defined Successive Difference Replication (SDR) variance estimators to generalize either 
of two quadratic-form estimators of variance of $\hat{Y}$ presented in Wolter (1984), 
\beq  \mbox{\bf SD} \, = \, \frac{1}{2}\,  \sum_{j=1}^{n-1} \, (\breve{y}_ j^{\circ}-  \breve{y}_{j+1}^{\circ})^2 \quad, 
\qquad \mbox{\bf SD2} \, = \, \frac{1}{2} \,  (\breve{y}_ 1^{\circ} -  \breve{y}_n^{\circ})^2  + \sum_{j=1}^{n-1} \, 
(\breve{y}_ j^{\circ} -   \breve{y}_{j+1}^{\circ})^2    \label{SDs}  \eeq 
In this paper we discuss only the SDR method based on SD2, as that is the one implemented in ACS and other 
Census Bureau surveys. 
\par Fay and Train's (1995) SDR variance estimator is based upon a fixed $R \times R$ Hadamard matrix ${\bf H}$, 
where the entries $h_{t,r}$ of ${\bf H}$ are  numbers $\pm 1$ with orthogonal rows such that $\sum_{r=1}^R \, h_{t,r}
 = 0$ for all $t \ge 2$. The matrix ${\bf H}$  used for this purpose, with $R$ a multiple of $4$, is most often chosen 
according to a method of Paley or Plackett and Burman (1946), implemented in the {\tt R} library {\tt survey} 
in functions {\tt paley} and {\tt Hadamard}. The entries $h_{t,r}$ are used to define multiplicative 
weight-factors, centered at $1$, with respect to a specific ordering $\ul{\bf a} = (a_1, \, a_2, \, \ldots, \, a_m)$ 
(without repeats, i.e., a permutation) of $m \le R-1$ elements of $\{1, \ldots, R\}$ according to the rule
\beq f_{j,r}({\bf a}) \, \equiv \, f_{j,r}= 1 \, + \, 2^{-3/2} \, (h_{a_k,r} - h_{a_{k+1},r}) \;\;, \quad 1 \le 
j \le n-1, \quad k  -1 =  (j-1) \mod \, m  \label{fdef} \eeq
where $a_{m+1} \equiv a_1$ by notational convention, and for $k - 1 = (n-1) \mod \, m$,
$$f_{n,r} \equiv \left\{ \ba{l@{\quad {\rm if} \quad}l}  1 + 2^{-3/2} (h_{a_k,r} - h_{a_1,r}) & 
n \le m \vspace{1.5mm} \\   1 + 2^{-3/2} (h_{a_k,r} - h_{a_{k+1},r} )  &
 n > m \ea \right.$$
This definition of replicate weight-factors applies to an arbitrarily large number $n$ of indices $j$ 
based on a single cycle $\ul{\bf a}$ of $m$ permuted indices of $\{2,\ldots, R\}$.  Section~\ref{cycles} gives 
a more general detailed definition based on $D > 1$ successive cycles, a method in general use in the Census 
Bureau and elsewhere.

We complete our basic definition of SDR with the formula for the variance estimator. For each replicate index 
$r=1,\ldots, R$, the $r$'th replicate total estimator is 
$$ \hat{Y}_r \; = \; \sum_{j=1}^n \, f_{j,r} \, \breve{y}_j^{\circ} $$
and in terms of the replicate estimators, \vspace{-2mm}
\beq \hat{V}^{\tt SDR}(\hat{Y}) \; = \; \frac{4}{R} \, \sum_{r=1}^R \, (\hat{Y}_r \, - \, \hat{Y})^2 \label{Vdef} \eeq
When $n \le m$, Fay and Train (1995) and later authors showed that the variance estimator (\ref{Vdef}) is 
algebraically equivalent to {\bf SD2} in (\ref{SDs}). This fact also follows by expressing SDR as a group-level 
variance-estimation method in Section~\ref{dist.gps}.

\subsection{Multiple-Cycle Definitions of SDR}  \label{cycles}

Although several published studies of SDR and its properties (Huang and Bell 2009, 2010; Opsomer et al.~2016) 
restrict themselves to definition (\ref{fdef}) based on a single cycle of permuted indices, SDR has been implemented 
in ACS and in CPS (for replicate weighting in Self-Representing strata) by a method involving $D > 1$ permutation 
cycles {\bf a} for indexing row-pairs from a single $R \times R$ Hadamard matrix. The idea, already suggested by 
Fay and Train (1995), aims to produce many more weighting-groups than replicates. We define the general method in this 
Section and later recommend values of $R, D$ based on the statistical properties of the resulting SDR variance estimators.

Variant methods of extending (\ref{fdef}) based on $D>1$ permutation cycles were presented by Navarro (2001), 
Sukasih and Jang (2003), and Ash (2014). All were similar and can be discussed together using the following notation. 
With $R$  a multiple of $4$ and $m = R-1$ or $R-2$, fix a subset ${\cal B}$ of $m$ indices in $\{1,\ldots, R\}$. 
(Generally, ${\cal B}$ has been taken to be $\{1, \ldots, R/2, R/2+2,\ldots, R\}$ or $\{2, \ldots, R/2, R/2+2,\ldots, R\}$.)  
Then for $d=1,\ldots, D$, let $\mathbf{a}^{(d)} = (a^{(d)}_k, \; k=1,\dots, m)$ denote a permutation of ${\cal B}$, written 
as an ordered $m$-tuple.
However they are generated, the permutations $\mathbf{a}^{(d)}$ are assumed to be cyclically re-ordered so that for 
all $d \ge 1, \;\; a^{(d+1)}_1 = a^{(1)}_1$, and the successive permutations can then be concatenated into a single long 
string $(\gamma_1, \ldots, \gamma_{Dm})$, where  $~\gamma_{dm+1} \, = \, \gamma_1$~ for $d=1,\ldots, D-1~$ and
\beq \gamma_{(d-1)m+k} \; = \; a_k^{(d)}, \qquad \mbox{for} \quad k=1, \ldots, m, \quad d=1,\ldots D \label{long.cycle} \eeq
and by  convention define ~$\gamma_{Dm+1} \equiv \gamma_1$. In terms of this indexing, define 
\beq f_{j',r} \; = \;  1 \, + \, 2^{-3/2} \, (h_{\gamma_j,r} - h_{\gamma_{j+1},r}) \;\;, \quad \mbox{for} 
\quad 1 \le j' \le n-1, \;\;\; j \, = \,  j' \mod Dm  \label{fdefD} \eeq
except in the case where $n < m$: in that case, for all $D$,
$$ f_{n,r} \; \equiv \;   1 + 2^{-3/2} (h_{\gamma_n,r} - h_{\gamma_1,r}) $$
When $n \le m$, this definition of $\{f_{j,r}\}$ is identical to (\ref{fdef}) and wraps around at $j=n$ to reproduce SD2.  
For larger $n$, indexing in (\ref{fdefD}) is periodic along indices $j=1,\ldots, n$ with period $~Dm$, 
associating index $j = (d-1)m + k$ with the row-index pair  $(\gamma_j, \gamma_{j+1}) = (a_k^{(d)}, a_{k+1}^{(d)})$ 
of the Hadamard matrix ${\bf H}$. This row-indexing is done $D$ times successively using permutations $\mathbf{a}^{(d)}$ in place of $\mathbf{a}$, with $d=1,\ldots, D$, and then repeated for indices $Dm < j \le n$. To enable $~f_{j,r} = 1 + 2^{-3/2} (h_{\gamma_j,r} - h_{\gamma_{j+1},r})~$ to be the unified definition of replicate weight-factors for all $j \le n$ and all $D$, we define $\gamma_j$ to be equal to $\gamma_1$ if $j=n+1 \le m$ and otherwise, for all $j > Dm$, to be periodically continued from $j=1,\ldots, Dm$.

For application to ACS with $R=80, D=10$, all authors including Navarro (2001) and Ash (2014) use the same 
verbal description of a row-indexing scheme $\mathbf{a}^{(d)}, d=1,\ldots,D$, as in Sukasih and Jang (2003) 
with successive cycles modified from arithmetic progressions with increasing spans, but it is not clear in any of the papers except Navarro (2001) -- where SAS code is given -- exactly how the $10$-cycle indexing was done. Two different implementations, with slightly different numbers of duplicated replicates, are described in greater detail in Appendix~\ref{cycle10} and are reproduced explicitly in Supplement~\ref{ACScycles}.

In Fay and Train's (1995) multi-cycle indexing, elaborated by later authors, the successive set of $Dm$ row-index pairs \ $(\gamma_k, \, \gamma_{k+1})$ \ should avoid duplicating one another. However, the schemes described by Navarro (2001) and Sukasih and Jang (2003) do generate a nonempty but relatively small set of duplicate pairs, with details in Appendix~\ref{cycle10} and Supplement~S1. Every scheme with $m=78$ and $D=10$ that we know of produces some duplicate pairs. Appendix~\ref{cycle10} and Supplement~S1 exhibit a maximal subset of $7$ cycles of length $78$ without duplicates within the $10$ cycles constructed by the SAS code in Navarro (2001). The general problem of characterizing the largest number $D$ of cycles possible for a given $m$ which produces no duplicated row-assignment pairs may be of mathematical interest, but we show in Sections~\ref{dist.gps} and \ref{SDR.Bias} why it is not very important statistically.

\subsection{Objectives and Organization of the Paper} \label{organiz}

The quality of SDR variance estimators should be evaluated both by their bias and variance regarded as 
design-based survey-sample estimators. Past evaluations of SDR variance estimators have generally been based on simulations  focused on the bias. An exception is the paper of Huang and Bell (2009) which also presented theory and simulations  approximating the SDR estimator as approximately a chi-squared distributed random variable. The objectives of the present paper are: \vspace{-3mm}
\bi   \item[(a)] to provide simple computing formulas for SDR variance estimators of totals, without Hadamard matrices, for general $D$ and $R$, and to relate them to {\bf SD2} formulas for weighted group totals; \vspace{-1mm}
\item[(b)] to extend Huang and Bell's (2009) discussion of SDR variance to more complex sampling designs \linebreak and multiple-cycle indexing,  and to recommend numbers $R$ of Hadamard-matrix columns \linebreak and $D$ of cycles; \vspace{-1mm}
\item[(c)] to develop theoretical formulas for bias of SDR variance-estimators, in the ideal setting of stratified 
SRS samples from stratumwise {\tt iid} superpopulations; \vspace{-1mm}
\item[(d)] to use the formulas (c) and simulations to assess the biases of SDR variance estimators 
for survey-weighted estimates on domains, with a view to application in small-area estimation; and \vspace{-1mm}
\item[(e)] to discuss the relevance for small-area estimation of bias and variance of SDR variance estimators. \ei\vspace{-1mm}

The rest of the paper is organized as follows. Section~\ref{history} summarizes the previous history of SDR evaluations. Section~\ref{dist.gps} provides theory on the distinct weighting-groups arising with $D\ge1$ and applies it to variances of the SDR estimators. Section~\ref{bias.theory} develops superpopulation theoretical formulas for the biases of SDR estimators of variance on domains. These formulas lead in Section~\ref{SDR.Bias} -- through study of special cases of the formulas and simulations of a variety of parameter-defined populations and sampling designs -- to a general picture of the relative biases of SDR variance-estimators and what features of samples and domains they depend on. Section~\ref{Disc} draws conclusions for Small Area Estimation (SAE) projects, especially those based on many small domains.

\section{Previous Simulation Studies of SDR} \label{history}

Because of the importance of SDR variance estimates from the ACS and CPS, previous authors have published simulation studies of the validity of these variance estimates. These studies have all used random finite superpopulations, mostly {\tt iid}, with only Sukasih and Jang (2003) employing unequal probability weights, and have considered only statistics (usually totals) based on the entire sample. These previous studies differ in their choice between two ({\bf SD2} or {\bf SD}) versions of SDR indexing: Sukasih and Jang (2003) and Ash (2014) like us adopt the {\bf SDR2} version, while Huang and Bell (2009) and Opsomer et al.~(2016) use {\bf SD}.

Sukasih and Jang  (2003) studied the behavior of some nonlinear estimates (ratio estimators, correlation 
and regression coefficients, and medians) from the National Survey of College Graduates (NSCG) and from 
simulations of stratumwise simple random (SRS) samples of size 2 per stratum from various finite 
populations with 32 strata and two attributes with stratumwise bivariate-normal distributions. They found 
that the SDR variances for totals behaved `equivalently' -- and for the nonlinear statistics other than medians, 
`comparably' --  to Jackknife or Taylor Series estimates in their computed examples. For medians, they 
found that SDR and Taylor series estimates remained comparable, while the jackknife method produced 
positive bias and unstable estimates.

Huang and Bell (2009) studied the distributions of whole-population SDR variance estimates of totals 
from equally weighted  {\tt iid} superpopulations with differing underlying attribute distributions (Normal, 
Bernoulli and Poisson) and sample sizes (from 2 to 760).  Their results for $n \ge 100$  were `mostly supportive' 
of using a chi-square distribution for the SDR variance estimator, with degrees of freedom varying considerably 
across the different choices of underlying attribute distribution. For small sample sizes (less than 100), they found 
little evidence of bias in SDR, with serious biases arising only `for a very few cases with very small sample sizes
 [$n < 15$]'.  Further simulations using (SRS samples from) an artificial population of ACS 2005 data gave similar 
conclusions. But they did not simulate samples with unequal weights. Huang and Bell (2010) continued their 
simulations regarding SDR by comparison with other replication methods and some nonlinear statistics of 
sampled domains. 

Ash (2014) described and studied multi-cycle SDR estimates. His simulations used SRS samples of 7 artificial random superpopulations of size $N=64,000$ following Wolter (1984), from each of which a sample of 64 units was drawn systematically. In that setting, only 100 samples were possible from each  superpopulation, and SDR variance estimates for totals were calculated with each of R=16, 32, 48 and 64 replicates. Larger numbers of replicates were found to have little effect on the bias (except for some estimators in superpopulations with linear trend) but an expected systematic decreasing effect on variance of the variance estimator.

Finally, the simulation study of Opsomer et al.~(2016) drew systematic samples of size $n=100$ from 
random superpopulations of size $N=2000$ in which the attributes $y_i$  were correlated with the variable 
$z_i$ used to perform the sort-ordering of the sample. Quite different from the previous studies, these 
simulations studied aspects of SDR performance: the correlation and linear or nonlinear dependence 
attributes $y_i$ and the sorting variable; the single-stage systematic sampling design versus a two-phase 
sampling design; and also the alteration of the coefficient $2^{-3/2}$ within the definition  (\ref{fdef}) to 
$(1-\epsilon)^{3/2}$, also changing the factor $4$ in (\ref{Vdef}) to $(1-\epsilon)^{-2}$. These 
simulations generally supported the adequacy of SDR variance estimators in systematic sampling, with the 
standard value $\epsilon=0.5$ as good as any, and superiority to a BRR method used for comparison. In 
the two-phase setting, especially in simulations based on NSCG data, some variance over-estimation occurred 
but could largely be corrected by ratio-adjustment of replicates with respect to first-phase frame totals.

Of these previous simulation studies, all except those of Opsomer et al.~(2016) involved multiple (1000 or
more) samples drawn from each randomly generated superpopulation. The Opsomer et al.~(2016) 
simulations generated a new superpopulation and a systematic sample (one of 20 possible such samples) 
from it in each simulation iteration.

\section{Distinct Weight-Replication Groups and SDR Variances} \label{dist.gps}

Let $n$ be arbitrarily large   
and  $\{f_{j,r}\}$ be a system of replicate weight-multipliers, as defined in (\ref{fdefD}), based on a full-rank $R \times R$ Hadamard matrix ${\bf H}$. Then the SDR variance estimator (\ref{Vdef}) in terms of replicate estimators 
$\hat{Y}_r$ can be rewritten as
$$ \hat{V}^{\tt SDR}(\hat{Y}) \; = \; \frac{4}{R} \, \sum_{r=1}^R \, \Big(\sum_{j=1}^n \, (f_{j,r}-1) \, 
\breve{y}_j^{\circ} \Big)^2 \; = \; \sum_{j=1}^n \sum_{k=1}^n \, \Big[\frac{4}{R} \, \sum_{r=1}^R \,  
(f_{j,r}-1)(f_{k,r}-1) \Big] \, \breve{y}_j^{\circ} \, \breve{y}_k^{\circ}$$
Now enumerate the {\it distinct\/} vectors  $\mathbf{f}_j \, \equiv \, (f_{j,r}, \; r=1,\ldots, R) \in \mathbb{R}^R$ of 
weight-multiplier replicates as  $\mathbf{v}_g, \; g=1,\ldots, G$,  where $\mathbf{v}_g = (v_{g,r}, \; 
r=1,\ldots,R) \in \mathbb{R}^R$, and define groupwise weighted totals 
\beq  Y^{(g)} \; \equiv \; \sum_{j=1}^n \; I_{[ \mathbf{f}_j = \mathbf{v}_g]} \;  \breve{y}_j^{\circ}
\label{Ygdef} \eeq
In the case where $n \le m$, regardless of $D$, each $Y^{(g)} =   \breve{y}_g^{\circ}$ for $g\le n$. When $D=1$ and  $n>m$, the group totals are still explicit and simple, defined by arithmetic progressions of indices,
$$ Y^{(g)} \; = \; \sum_{k \ge 0} \; I_{[km + g \le n]} \;  \breve{y}_{km+g}^{\circ} \qquad \mbox{for} \quad g=1,\ldots, m$$ 
Under the restriction that the cycles $a_k^{(d)}$ and $\gamma_j$ have been chosen in such a way that all ordered pairs $(\gamma_j,\gamma_{j+1})$ are distinct for all $j \le n$, then for $m < n \le Dm$, again the group-totals $Y^{(g)}$ are 
identical to singleton $\breve{y}_g^{\circ}$ values, and the groups are defined by arithmetic progressions of indices 
for $D > 1, n >Dm$.

By (\ref{Ygdef}), for all $D, n$, the SDR variance is a quadratic form in the group totals, 
\beq \hat{V}^{\tt SDR}(\hat{Y}) \; = \; \sum_{g=1}^G \, \sum_{g'=1}^G \, Q_{g,g'} \, Y^{(g)} \, 
Y^{(g')} \;, \qquad Q_{g,g'} \; = \; \frac{4}{R} \, \sum_{r=1}^R \,  (v_{g,r}-1)(v_{g',r}-1) \label{VsdrG} \eeq
where $Q$ has diagonal elements $1$ and  therefore $tr(Q) = G$. (This follows because for each 
$g$, there is a pair of distinct indices $j, j^{\ast}$ for which $v_{g,r} - 1 = 2^{-3/2}(h_{j,r}-h_{j^{\ast},r})$ 
for all $r$.)

The quadratic form representation (\ref{VsdrG}) relates SDR to so-called `random groups' methods of variance estimation (Wolter, 2007) where sample groups $\{j: \,  \mathbf{f}_j  = \mathbf{v}_g\}$ are partially random due to the sort-ordering of the respondent sample, generally in the same order in which sampling (often systematic at the final stage) was done. In practice, the sort ordering may be complicated. In the implementations of SDR in ACS, the geographic sort-ordering used is not identical to the geographic sort from which ACS sampling was done, but is based on a re-sort taking both first- and second-stage sampling into account. In CPS, the sort-ordering is basically geographic, but modified based on month of sampling.  

When the weight multipliers $f_{j,r}$ are defined as in (\ref{fdef}), with $D=1$, the number of distinct weight-multiplier 
vectors $\mathbf{f}_j$ is ~$\min(m,n)$, in one-to-one correspondence with the pairs $(a_k, a_{k+1})$ 
of row-numbers for $j=1,\ldots, \min(m,n)$ (with $(a_n,a_1)$ in place of $(a_n, a_{n+1})$ when $n < m$). In this case, $G= \min(m,n)$ and the matrix $Q$ is explicitly calculated and does not depend on the specific  Hadamard matrix ${\bf H}$ used in (\ref{fdef}), since for $g,g' = 1, \ldots, m$, the orthogonality of distinct rows of ${\bf H}$ implies  
$$ Q_{g,g'} \; = \; \frac{4}{R} \, \sum_{r=1}^R \, 2^{-3} \, (h_{a_g,r}-h_{a_{g+1},r}) \,
  (h_{a_{g'},r}-h_{a_{g'+1},r})  $$  
$$ \quad = \frac{1}{2R} \, \sum_{r=1}^R \,  (I_{[g=g']} \, + \, I_{[g = g'\pm 1 \\ \mod m]}) \, (h_{a_g,r}-h_{a_{g+1},r})\,
  (h_{a_{g'},r}-h_{a_{g'+1},r})$$
\beq  \qquad  = \; I_{[g = g' \!\!\mod m]} \, - \, \frac{1}{2} \, I_{[g= g' \pm 1 \!\! \mod m]} \label{Qsdr2} \eeq 
Thus, when weight-factors are based on a single cycle $\{a_k\}_{k=1}^m$ (the case $D=1$ of Section~\ref{cycles}), equation (\ref{VsdrG}) yields $Q$ as a tridiagonal $G \times G$ matrix (with $G = \min(m,n)$) with entries $1$ on the main diagonal and $-1/2$ 
on the above-and-below diagonals, and the SDR variance as
\beq \hat{V}^{\tt SDR} \; = \; \sum_{g=1}^G \Big\{ (Y^{(g)})^2 - Y^{(g)} Y^{(g+1)}\Big\} \; = \; 
\frac{1}{2} \, \sum_{g=1}^G \, (Y^{(g)} - Y^{(g+1)})^2 \label{SDGp} \eeq
and by notational convention, $Y^{(G+1)} \equiv Y^{(1)}$.  However, $Q$ and $\hat{V}^{\tt SDR}$ are not so simple when $D>1$.

Consider the general case $D\ge 1$ for any $n > 1$, in terms of the single long sequence $\{\gamma_j\}_{j=1}^{n+1}$ with $\gamma_{km+1}$ identical for all $k=0, \ldots, D-1$, as in (\ref{long.cycle}), periodically contnued from indices $1,\ldots, Dm$ when $n > Dm$, and with $\gamma_{n+1} = \gamma_1$ when $n < m$. By the nonsingularity of ${\bf H}$, replicate vectors $f_{j,r} = 1 + 2^{-3/2} (h_{\gamma_j,r} - h_{\gamma_{j+1},r})$ corresponding to  $(\gamma_j, \gamma_{j+1})$  and $f_{j',r}$ corresponding to $(\gamma_{j'}, \gamma_{j'+1})$ are exactly identical 
as vectors indexed by $r=1,\ldots, R$ if and only if the row-index pairs $(\gamma_j, \gamma_{j+1})$  
and $(\gamma_{j'}, \gamma_{j'+1})$ are equal.  Indices $g=1, \ldots, G$ enumerate the set of distinct row-index pairs $(\gamma_j, \gamma_{j+1})$, with $v_{g,r} \equiv f_{j,r}$ enumerating the corresponding vectors (indexed by $r=1,\ldots,R$) of replicate-weight-factors.  Here $G \le \min(n, Dm)$, and if $g, g'$ are indices such that $v_{g,r} = f_{j,r}, \; v_{g',r} = f_{j',r}$, then 
\bea Q_{g,g'} & = & 
        I_{[\gamma_j=\gamma_{j'}, \, \gamma_{j+1}=\gamma_{j'+1} ]} \; - \; I_{[\gamma_j=\gamma_{j'+1}, \, \gamma_{j+1}=\gamma_{j'} ]}\; + \; \frac{1}{2}\, I_{[\gamma_j =\gamma_{j'}, \, \gamma_{j+1} \ne \gamma_{j'+1}]} 
 \nonumber \\
& + &  \frac{1}{2}\, I_{[\gamma_j \ne \gamma_{j'}, \, \gamma_{j+1} = \gamma_{j'+1}]}  \; - \; 
\frac{1}{2} \,  I_{[\gamma_j = \gamma_{j'+1}, \, \gamma_{j+1} \ne \gamma_{j'}]} \; - \; 
\frac{1}{2} \,  I_{[\gamma_j \ne \gamma_{j'+1}, \, \gamma_{j+1} = \gamma_{j'}]} 
\nonumber  \vspace{1.5mm} \\
 & = &  \frac{1}{2}\, \Big[ I_{[\gamma_j = \gamma_{j'}]} \, + \, I_{[\gamma_{j+1} = \gamma_{j'+1}]}  \, - \,
I_{[\gamma_j = \gamma_{j'+1}]} \, -  \, I_{[ \gamma_{j+1} = \gamma_{j'}]} \Big] 
 \label{Qalt.D} \eea
Define the $n \times n$ matrix $M$ with entry $M_{j,j'}$ given by the right-hand side of (\ref{Qalt.D}). Then the definition of $\hat{V}(\hat{Y})$ and formula (\ref{VsdrG}) imply also
\beq \hat{V}^{\tt SDR}(\hat{Y}) \, = \, \sum_{j=1}^n \, \sum_{j'=1}^n \, M_{j,j'} \, \breve{y}_j^\circ \, \breve{y}_{j'}^\circ 
\label{VSDR.M} \eeq

The formulas of this Section explicitly enable the calculation of SDR estimates of survey-weighted estimates $\hat{Y}$  of population totals $(\hat{\ul{Y}}$ if vector-valued), without the need to retain the SDR replicates $\hat{Y}_r$, thus fulfilling objective (a) of Sec.~\ref{organiz}. The formulas, valid for all $n$, are (\ref{VsdrG}) coupled with (\ref{Qsdr2}) for general $D$, with the simpler formula (\ref{SDGp}) valid for $D=1$.  The occasions when replicate weight factors \emph{are} needed, as explained in Krewski and Rao (1981), is in  constructing implicitly linearized estimates of variances of nonlinear functions $g(\hat{\ul{Y}})$ (such as ratio-estimators) of survey-weighted vector attributes totals $\hat{\ul{Y}}$, in which case the variance estimates are ~$(4/R) \, \sum_{r=1}^R \, (g(\hat{\ul{Y}}_r) - g(\hat{\ul{Y}}))^2$.

SDR variance estimation is generally upwardly biased for small domains. The explanation is simple, at least for $D=1$, when formula (\ref{SDGp}) expresses $\hat{V}^{\tt SDR}$ as one-half of a sum of squared group differences. If the weighted group totals $Y^{(g)}$ are {\tt iid} (which might be true if the $\breve{y}_i$ are {\tt iid} and $n$ is a multiple of $m$), and if the sizes of sampled groups $g$ are conditioned on being positive, the expectation of $(Y^{(g)}-Y^{(g+1)})^2/2$ is the variance of $Y^{(g)}$. However, 
when many of the sample sizes are small and many of the groups are empty, the vector $(Y^{(t)}, \; 
t=1,\ldots, G)$ of group totals contains many $0$'s with high probability, and the unconditional expectation of 
terms $(Y^{(g)}-Y^{(g+1)})^2/2$ may be much larger than the variance of $Y^{(g)}$. When $D>1$, such 
a clear explanation is not possible, although we will find similar upward small-domain biases in SDR variance 
estimates, by theoretical formulas and simulation.

\subsection{Variances of the SDR Estimates}  \label{SDR.Variance} 

Validity of the SDR approximation to true variance is sometimes presented in terms of the similarity of 
neighboring attributes $y_i$ in the sample sort-order defined for SDR. Taking sample weights into account 
(after such operations as  nonresponse adjustments like raking or calibration or simply because of 
nonresponse propensity modeling which renders weights stochastic), this could be amended to require 
distributional similarity of neighboring weighted attributes $\breve{y}_j = y_j \, w_j$. On the other hand, the representation (\ref{VsdrG}) of $\hat{V}^{\tt SDR}$ as a quadratic form in group subtotals $Y^{(g)}$ suggests 
that triangular-array superpopulation assumptions supporting validity of SDR should result in approximate 
restrictions on distributional behavior at group level $Y^{(g)}$, more or less as in BRR (Krewski and Rao 1981,
 Wolter 2007). For larger groups, design-based  central limit theorems instead of model-based superpopulation 
assumptions might be used to justify approximate normality of the design sampling distributions of weighted 
group totals $Y^{(g)}$. Justification of the SDR variance as a good approximation to the true variance of 
$\hat{Y}$ can be given by assuming approximate {\tt iid} behavior of the group totals $Y^{(g)}$ as random 
variables. Under assumptions making the $Y^{(g)}$ variables numerous or approximately normal, the quadratic 
form (\ref{VsdrG}) becomes approximately a (nonrandomly) weighted sum of chi-squared random variables, 
which itself can be approximated as a chi-squared random variable. This was the motivation for the chi-squared 
approximations of Bell and Huang (2009) to the sampling distribution of $\hat{V}^{\tt SDR}$.

 Starting from formula (\ref{VsdrG}) or (\ref{VSDR.M}), the variance of the estimator $\hat{V}^{\tt SDR}$ can be computed assuming that the population values $y_i$ are {\tt iid\/} and that sampling is SRS. Formula (\ref{VsdrG}) provides a way to calculate $ var(\hat{V}^{\tt SDR})$ as in Huang and Bell (2009), in terms of the spectral decomposition of the matrix $Q$, but for general $n, \, D$ the group subtotals $Y^{(g)}$ are not {\tt iid} as they were for $D=1$ and $n$ a multiple of $m$. So  (\ref{VSDR.M}) leads to a more convenient general formula (\ref{VSDRvar}) for $var(\hat{V}^{\tt SDR})$ which is given in Appendix~\ref{App2.Var}  in terms of the first four moments of $y_i$. This formula, implemented in an {\tt R} function {\tt VarSDRVar}, can be used to exhibit the dependence of the variance on the cycling number $D$, for fixed $n, m$. Table~\ref{vartab1} displays the ratio of $\{ var(\hat{V}^{\tt SDR}) \}^{1/2}$ over the target $var(\hat{Y})$ of estimation, when $n$ is an integer multiple $K$ of $m$, and $\mu_4 = 3 \sigma^2$, and the sampling fraction $n/N$ can be ignored. 
The entries of  Table~\ref{vartab1} are roughly the coefficients of variation (CVs) of $\hat{V}^{\tt SDR}$ for the indicated combinations of $K$ and $D$.

\begin{table} \caption{Ratio of $\{var(\hat{V}^{\tt SDR})\}^{1/2}$ over $var(\hat{Y})$, when $n/N$ is negligible, for $y_i$ iid with kurtosis 3, based on data SRS sampled, with $m=78, n = K\cdot m$. } \label{vartab1}
\begin{center}
\begin{tabular}{r|rccccccccc}
K  &   D= 1   & 2  & 3  & 4  & 5  & 6 &  7 &  8 &  9 & 10  \\ \hline
1  &         0.196 & 0.196 & 0.196 & 0.196 & 0.196 & 0.196 & 0.196 & 0.196  & 0.196 & 0.196 \\
2  &         0.196 & 0.179 & 0.179 & 0.179 & 0.179 & 0.179 & 0.179 & 0.179  & 0.179 & 0.179 \\
3  &         0.196 & 0.181 & 0.173 & 0.173 & 0.173 & 0.173 & 0.173 & 0.173 & 0.173 & 0.173  \\
4  &         0.196 & 0.179 & 0.175 & 0.170 & 0.170 & 0.170 & 0.170 & 0.170 & 0.170 & 0.170  \\
5  &         0.196 & 0.180 & 0.174 & 0.171 & 0.168 & 0.168 & 0.168 & 0.168 & 0.168 & 0.168  \\
6  &         0.196 & 0.179 & 0.173 & 0.171 & 0.169 & 0.167 & 0.167 & 0.167 & 0.167 & 0.167  \\
7  &         0.196 & 0.180 & 0.174 & 0.171 & 0.169 & 0.168 & 0.166 & 0.166 & 0.166 & 0.166  \\
8  &         0.196 & 0.179 & 0.174 & 0.170 & 0.169 & 0.168 & 0.166 & 0.165 & 0.165 & 0.165  \\
9  &         0.196 & 0.179 & 0.173 & 0.170 & 0.169 & 0.168 & 0.167 & 0.166 & 0.165 & 0.165 \\
10 &        0.196 & 0.179 & 0.173 & 0.170 & 0.168 & 0.167 & 0.167 & 0.166 & 0.165 & 0.164  \\
\hline \end{tabular} \end{center} \end{table}

This Table already indicates that for $n$ an integer multiple of at most $10$ times $m$, the CV of the SDR variance depends only weakly on $D$ as long as $D$ is at least 3 or 4. As remarked in Appendix~\ref{App2.Var}, the definition of SDR implies that for fixed $n, m$ the estimate $\hat{V}^{\tt SDR}$ is the same for all choices of $D$ at least $\ge n/m$ and that the approximate CV ratios $acv = \{var(\hat{V}^{\tt SDR})\}^{1/2}/var(\hat{Y})$ do not vary with $\mu, \mu_3, \sigma^2$ when $n$ is a multiple of $m$. On the other hand, when $n$ is not a multiple of $m$ these ratios do depend on $D, \mu, \mu_3$ and $\sigma^2$. For example, for $n$ ranging from $10$ to $77$, the ratio $acv$ does not depend at all on $D$ and, for $\mu/\sigma = \mu_3/\sigma^3=1$, decreases monotonically in $n$ from $0.548$ down to $0.197$. For $n$ in the  range $79$ to $146$, with $D=1$ the ratio $acv$ for $m=78, \, \mu_4=3\sigma^4$ and large $N$ rises from $0.203$ up to $0.209$ (near $n=117$) and then decreases back to $0.199$. When the same $acv$ ratios are computed using $D=2$ for $79 \le n \le 155$, they are smaller for $D=2$ by an amount that increases monotonically from about 1\% to about 8\% as compared with $D=1$ as $n$ ranges from $79$ to $153$. Thus, as in Table~\ref{vartab1}, the variance of $\hat{V}^{\tt SDR}$ is smaller for $D=2$ for these $n$ values; but the variances are essentially as small as possible as long as $D$ is as large as $3$ or $4$.

Table~\ref{vartab1} shows that the CV of the SDR variance estimate $\hat{V}^{\tt SDR}$ does not decrease with $n$. As was stated in the paper of Krewski and Rao (1981) and the chapters of Wolter (2007) on random groups and replication-based variance estimation, the precision of variance estimates based on replicates depends primarily on the number of replicates ($R=m+2$ in our case, usually equal to $80$) rather than on $n$. The CV value roughly $0.2$ is not untenably large for $n=78$ or $156$), and for larger values of $n$ (equal to $78 K$ for $K\ge 3)$ the standard deviation of $\hat{V}^{\tt SDR}$ in Table~\ref{vartab1} is $\le \sqrt{0.2} \, \sigma^2 \, N/n^2$, small enough to be disregarded in many survey estimation problems.

These calculations of variance of SDR estimators suggest that $R$ should be as large as feasible, 80 in our examples, but the specific pattern of cycles chosen is not important nor is the number $D$ of cycles as long as it is at least 3 or 4. Ash (2014) had previously recommended using larger numbers of cycles in order that SDR be relatively unbiased for deterministically patterned (weighted) attributes.

\section{Theoretical Formula for SDR Bias in Stratified SRS Samples} \label{bias.theory}

The previous section presented variances of SDR estimators as a criterion for choosing $D, R$. However, in 
applications of SDR to domain estimation, the bias in variance estimation is at least as important. In the setting of 
stratified SRS designs, applied to a superpopulation that has identical means and variances for all units within strata, 
a theoretical formula for the bias and relative bias can be derived. This formula will be used to assess the relative bias 
in SDR estimation and the characteristics of the superpopulation and domains with a noticeable effect on bias.

Consider a finite population ${\cal U}$ partitioned into a collection of population {\it strata\/} ~${\cal U}_h$, 
$h=1,\ldots, H$, and a subset $A$ which is the {\it domain\/} on which the total of attributes $y_i$ is of interest. Let $~\epsilon_i \, \equiv \, I_{[i \in A]}$,~ and let $h(i)$ denote the stratum label $h$ for which $i \in {\cal U}_h$, \ and 
\vspace{-1mm}
\beq  |{\cal U}| \, = \, N \; , \quad |{\cal U}_h| \, = \, N_h \; , \quad |A| =\, N_A \; , \quad |A \cap
 {\cal U}_h| \, = \, N_{Ah} \; , \quad  p_h \, = \, \frac{N_h}{N} \; ,  \quad \alpha_h \, =  \frac{N_{Ah}}{N_h} \label{N.not} \vspace{-1mm} \eeq
Assume about the finite population $~{\cal U}$ that \vspace{-1mm}
\begin{description} \item[{\bf (A.1)}]  The population ${\cal U}$ is ordered increasing with respect to 
stratum-index $h$, so that $~h \le h'$ whenever ~$i \in {\cal U}_h, \; i' \in {\cal U}_{h'}, \; i \le i'$. The 
attributes $y_i$ for $i \in {\cal U}$ are uncorrelated random variables with means $\mu_h$ and variances 
$\sigma^2_h$ that are constant over $i \in {\cal U}_h$, regardless of the membership $i \in A$.  \vspace{-1mm} \end{description}

\par The attribute of interest is $~z_i \, \equiv \, y_i \, \epsilon_i~$ in this Section. The domain total $~Y_A \equiv 
\sum_{i \in A} \, y_i \, = \, \sum_{i \in {\cal U}} \, z_i$ is to be estimated, based on a stratified SRS sample of size $n = \sum_{h=1}^H \, n_h$, ~of $n_h$ from ${\cal U}_h$, using 
\beq \hat{Y}_A \; = \; \sum_{h=1}^H \,  \sum_{i\in {\cal S}_h \cap A} \, (N_h/n_h) \, y_i  \; = \; \sum_{h=1}^H \, 
\sum_{j=n_1+\cdots+n_{h-1}+1}^{n_1+\cdots+n_h} \, \epsilon_j^{\circ} \, y_j^{\circ}/\pi_j^{\circ} \label{YhatA} \eeq
The unit single-inclusion probabilities for the sampled units are $\pi_j^{\circ} = n_{h^{\circ}(j)}/N_{h^{\circ}(j)}$, and the corresponding  survey weights are $w_j^{\circ} \, = \,  1/\pi_j^{\circ}$.

Standard design- and model-based SRS formulas for the variance $~V_{\tt tru}~$ of the estimator (\ref{YhatA}) are:
\bea \frac{1}{N^2}\, V_{\tt tru} & = & \frac{1}{N^2} \, \sum_{h=1}^H \, \frac{N_h (N_h-n_h)}{n_h} \, \frac{1}{N_h-1} \, 
\sum_{i \in {\cal U}_h} \, (z_i \, - \, \bar{z}_h)^2 \quad , \quad \mbox{where} \quad \bar{z}_h \; = \; \frac{1}{N_h} \, 
\sum_{i \in {\cal U}_h} \, z_i  \nonumber \\
&& \frac{1}{N^2} \, E(V_{\tt tru}) \; = \; \sum_{h=1}^H \; \frac{p_h^2}{n_h} \,(1 - \, \frac{n_h}{N_h}) \, \big\{\alpha_h 
\sigma_h^2 \, + \, \alpha_h(1-\alpha_h)\mu_h^2\big\} \; \equiv \; v_{\tt tru}  \label{Vtru} \eea
Under additional assumptions on the attributes $y_i$ being stratumwise {\tt iid}, the difference \linebreak $\big[ V_{\tt tru} \, - \, 
E(V_{\tt tru}) \big]/N^2$ converges to $0$ in probability as $N, n \to \infty$. However, the rest of this section presents nonasymptotic results comparing $E(V_{\tt tru})$ to $E(\hat{V}^{\tt SDR})$. All expectations in this Section are taken over SRS samples \emph{and} random attribute values $y_i$, conditioned on the numbers $N_{Ah}$ of superpopulation elements falling in strata partitioned by the domain $A$.

In this setting, assumption {\bf (A.1)}  enters explicitly through the following conditional distributional properties of sampled attributes in strata $h, h'$:
\bi
\item[{\bf (P.i)}] ~for all strata $h=1,\ldots,H$ and unit indices $i$,
$$P(\epsilon_i=1 \, | \, i \in {\cal S}_h) \; = \; \frac{N_{Ah}}{N_h} \; = \; \alpha_h \quad ,
 \qquad E(y_i^k\, | \, \epsilon_i=1, \, i \in {\cal S}_h) \; = \; \left\{ \ba{c@{\mbox{ \ if \ }}l} \mu_h & k=1 
    \vspace{1.5mm} \\  \mu^2_h+\sigma^2_h & k=2 \ea \right.$$
\item[{\bf (P.ii)}] ~for all  strata $h$  and distinct unit indices $i, \, i' \, \in \, {\cal U}_h$,
$$P(\epsilon_i=1, \, \epsilon_{i'}=1 \, | \, i, \, i' \in {\cal S}_h) \; = \; \frac{N_{Ah}\,(N_{Ah}-1)}{N_h \, (N_h-1)} \quad , 
\qquad E(y_i \, y_{i'} \, | \; \epsilon_i, \, \epsilon_{i'}, \;  i, \, i' \in {\cal S}_h) \; = \; \mu_h^2$$
\item[{\bf (P.iii)}] ~for all  pairs $h, \, h'$ of  distinct strata respectively containing unit indices $i, \, i'$,
$$P(\epsilon_i=1, \, \epsilon_{i'}=1 \, | \, i \in {\cal S}_h, \, i' \in {\cal S}_{h'}) = \alpha_h \,  \alpha_{h'} \;\; , \quad\;\; E(y_i \, y_{i'} \, | \, \epsilon_i, \, \epsilon_{i'}, \; i \in {\cal S}_h, i'\in {\cal S}_{h'}) = \mu_h  \, \mu_{h'}$$
\ei \vspace{1.5mm}

The starting point in developing a formula for the SDR variance and its expectation 
is formula (\ref{VsdrG}) coupled with (\ref{Qalt.D}),  
which says that 
\bea \hat{V}^{\tt SDR} & = & \sum_{j=1}^n \, \sum_{j'=1}^n  \,\breve{z}_j^{\circ} \, \breve{z}_{j'}^{\circ} 
\; \Big\{ I_{[\gamma_j=\gamma_{j'}, \, \gamma_{j+1}=\gamma_{j'+1} ]} \; - \; I_{[\gamma_j=\gamma_{j'+1}, \, \gamma_{j+1}=\gamma_{j'} ]}\; + \; 
\frac{1}{2}\, I_{[\gamma_j =\gamma_{j'}, \, \gamma_{j+1} \ne \gamma_{j'+1}]}  \nonumber \\
& + &  \frac{1}{2}\, I_{[\gamma_j \ne \gamma_{j'}, \, \gamma_{j+1} = \gamma_{j' +1}]}  \; - \; \frac{1}{2} \,  I_{[\gamma_j = \gamma_{j'+1}, \, \gamma_{j+1} \ne \gamma_{j'}]}
\; - \; \frac{1}{2} \,  I_{[\gamma_j \ne \gamma_{j'+1}, \, \gamma_{j+1} = \gamma_{j'}]} \Big\}  \label{Vsdr2} \eea
Taking account of the fact that the fifth and sixth indicators in the curly-bracketed expression in (\ref{Vsdr2}) are carried into one another by switching $j, j'$, we combine the terms  (within the double sum on $j, j'$)
$$  - \, \frac{1}{2} \, I_{[\gamma_j=\gamma_{j'+1}, \, \gamma_{j+1}\ne \gamma_{j'}]} \; , \;\;  - \, \frac{1}{2} \, I_{[\gamma_j \ne \gamma_{j'+1}, \, \gamma_{j+1} = \gamma_{j'}]} \quad \mbox{into} \quad - \, I_{[\gamma_j=\gamma_{j'+1}, \, \gamma_{j+1}\ne \gamma_{j'}]} $$
and then combine the terms
$$  - \, I_{[\gamma_j=\gamma_{j'+1}, \, \gamma_{j+1}\ne \gamma_{j'}]} \; , \;\; - \, I_{[\gamma_j=\gamma_{j'+1}, \, \gamma_{j+1}= \gamma_{j'}]} \quad \mbox{into} \quad - \, I_{[\gamma_j=\gamma_{j'+1}]} $$
The result is a slightly simplified formula
\bea \hat{V}^{\tt SDR} & = & \sum_{j=1}^n \, \sum_{j'=1}^n  \, \breve{z}_j^{\circ} \, \breve{z}_{j'}^{\circ} \; \Big\{   I_{[\gamma_j=\gamma_{j'}, \, \gamma_{j+1}=\gamma_{j'+1} ]} \; - \; I_{[\gamma_j=\gamma_{j'+1}]}\; 
\nonumber \\ & + & \frac{1}{2}\, I_{[\gamma_j =\gamma_{j'}, \, \gamma_{j+1} \ne \gamma_{j'+1}]}  
 \; + \; \frac{1}{2}\, I_{[\gamma_j \ne \gamma_{j'}, \, \gamma_{j+1} = \gamma_{j' +1}]} \, \Big\}  \label{Vsdr3} \eea

Coupling (\ref{Vsdr3}) with {\bf (P.i)}--{\bf (P.iii)} yields a computable formula for $E(\hat{V}^{\tt SDR})$ which is given for all $n, D$ in Appendix~\ref{D.relBias}. Under the further restriction that $n \le m$, for any $D$ all pairs $(\gamma_j, \, \gamma_{j+1})$ are distinct for $j \le n$, and that is used in Appendix~\ref{prop.proof} to prove a simpler and interpretable formula.

\begin{prop} Under Assumptions {\bf (A.1)}--{\bf (A.2)} and $n \le m$, \
$N^{-2} \; E(V^{\tt tru}) =v_{\tt tru}$  ~and 
\bea N^{-2} \, E(\hat{V}^{\tt SDR}) & = & v_{\tt tru} \, + \, \frac{1}{2}\, \sum_{h=1}^H \, \Big( \frac{\alpha_h \, p_h \, \mu_h}{n_h} \, - \, \frac{\alpha_{h+1} \, p_{h+1} \, \mu_{h+1}}{n_{h+1}} \Big)^2  \nonumber \\ & + &
\frac{1}{N} \, \sum_{h=1}^H \, p_h \Big[ \alpha_h \, \sigma_h^2 \, + \, \alpha_h \, (1-\alpha_h) \, \mu_h^2 \, \Big(1 \, + \; \frac{(n_h-1) \, N_h}{n_h^2 \, (N_h-1)} \Big) \Big] \label{VSDR2} \eea 
with $v_{\tt tru}$ given in (\ref{Vtru}). The second expression on the right-hand side of (\ref{VSDR2}) is negligible compared to $v_{\tt tru}$ when the sampling fractions $n_h/N_h$ are uniformly small. \label{Vform}\end{prop} \vspace{-3mm}

The practical conclusion is that $\hat{V}^{\tt SDR}$ is an upwardly biased estimator with superpopulation-model relative bias approximately equal to 
\beq \mbox{\bf RelBias} \; \approx \; \frac{1}{2} \;  \sum_{h=1}^H \, \Big( \frac{\alpha_h \, p_h \, \mu_h}{n_h} \, - \, \frac{\alpha_{h+1} \, p_{h+1} \, \mu_{h+1}}{n_{h+1}} \Big)^2  \, \Big/ \, \sum_{h=1}^H \; \frac{p_h^2}{n_h} \,(1 - \, \frac{n_h}{N_h}) \, \big\{\alpha_h \sigma_h^2 \, + \, \alpha_h(1-\alpha_h)\mu_h^2\big\}
\label{RelBias} \eeq

\begin{rem}  SDR has no automatic finite-population correction, although formula (\ref{Vtru}) does. Fay and Train and other authors suggest that such a multiplicative correction to SDR variances estimates should be used in some circumstances, although it is not clear when. The denominator $v_{\tt tru}$ in formula (\ref{RelBias})  shows why the correction could be needed, but also makes it hard to identify the correction to use when the sample fraction $n_h/N_h$ and domain fraction $\alpha_h$ vary dramatically across strata. \hfill $\Box$ \label{FPC} \end{rem}

\section{Examples and Simulations of SDR Bias} \label{SDR.Bias}

A few special cases of formula (\ref{RelBias}) help in interpreting the small-sample bias of SDR. Continue to assume a stratified SRS design with $n \le m \le R-1$ from a finite population satisfying condition (A.1), with $H \ge 2$, so that formulas (\ref{VSDR2}) and (\ref{RelBias}) hold, and assume for simplicity that the sampling fractions $n_h/N_h$ are small. To simplify notations, denote the domain fraction and expected domain sample size respectively by $p_A, \, n_A$.
Then
$$p_A \,= \, \frac{N_A}{N} \, = \, \frac{\sum_{h=1}^H \, N_{Ah}}{\sum_{h=1}^H \, N_h} \, = \, N^{-1} \sum_{h=1}^N \, \alpha_h \, N_h \;\; , \qquad n_A \, = \, \sum_{h=1}^H \, n_h \, \alpha_h$$

\noi {\bf Example (i).} If the domain belongs to only a single stratum (without loss of generality, stratum 1), so that $\alpha_h=0$ for $h>1$ and $p_A \, = \, \alpha_1, \; n_A = n_1 \alpha_1$, then
\beq  \mbox{RelBias} \; \approx  \;  \frac{p_A}{n_1 \, (\sigma_1^2/\mu_1^2+1-\alpha_1)}\; \le \; \frac{p_A^2}{n_A \, \sigma_1^2/\mu_1^2} \label{sngstrat} \eeq
is inversely proportional to the domain sample-size $n_A$, decreases with the squared coefficient of variation in stratum 1, and increases with the squared domain fraction $p_A^2$.

 In (\ref{RelBias}), alternating behavior among successive strata can lead to large SDR bias. We give two examples to make this clear.\footnote{Ash (2014) made a similar remark about alternation among successive sampled units, without explaining how that might arise in a sampling context. But here, alternation occurs among strata.} 

\noi {\bf Example (ii).} If $H$ is even, with $p_h, \, \alpha_h, \, \mu_h, \, \sigma_h^2$ constant across strata, while  $n_h$ strictly alternastes between two values, $~n_h = \, (n/H) \cdot \big( 1 + (-1)^h \, \rho)\big)$~ for a constant $\rho \in (0,1)$,~ then $p_A = \alpha_1$ and (\ref{RelBias}) specializes to
\beq \mbox{RelBias} \; \approx \;   H \cdot \, \frac{p_A}{n \, (\sigma_1^2/\mu_1^2 \, + \, 1-p_A)} \, \cdot \,  \big(\frac{2 \, \rho^2}{1-\rho^2}\big) \; = \; H \cdot \, \frac{p_A^2}{n_A \, (\sigma_1^2/\mu_1^2 \, + \, 1-p_A)} \, \big(\frac{2 \, \rho^2}{1-\rho^2}\big)  \label{alt.strat1} \eeq
Thus, in the case of strata with alternating sample size, the relative bias increases with the \linebreak degree of alternation and with the number of alternating strata, decreases with the squared coefficient of variation of attributes, and is inversely proportional to the domain sample size.

\noi {\bf Example (iii).} Assume $H$ is even and $n_h, p_h, \alpha_h, \sigma_h^2$ do not vary with $h$, but that $\mu_h \,= \, \mu \cdot \big( 1+ (-1)^h \rho\big)$.  Then  (\ref{RelBias}) gives
\beq \mbox{RelBias} \; \approx \;   H \cdot \, \frac{2 \, \rho^2 \, p_A^2}{n_A \, \big(\sigma_1^2/\mu^2 + (1-p_A)(1+\rho^2)\big)} \, \label{alt.strat2} \eeq

\noi {\bf Example (iv).} Formulas (\ref{VSDR2})--(\ref{RelBias}) scale systematically with respect to small $\alpha_h$ parameters. For a small domain within a larger survey (with moderate to large $n_h$ values across $h$), all $\alpha_h$ values may be small. Formulas  (\ref{VSDR2})--(\ref{RelBias}) show  that the numerator in the relative bias scales proportionately with the square of the average $\alpha_h$ value while  the denominator (\ref{Vtru}) scales proportionately with the average $\alpha_h$ apart from the factors $1-\alpha_h$ which are all close to $1$. This implies that SDR bias is small to negligible for domains which have uniformly small stratum fractions $\alpha_h$ within large stratified survey samples.

\noi {\bf Example (v).}  SDR variances can also be inflated when many sample units are isolated, i.e., domain sampled units $i$ occur in strata with neighboring strata containing no domain sample. Specifically, suppose that for some numbers $a_1, a_2, b > 0$ not varying with $h$,
$$\alpha_h \alpha_{h+1} = 0 \quad \mbox{for all} \;\; h \qquad \mbox{and} \quad a_1 \le \alpha_h/n_h \le a_2 \;\; , \quad \sigma_h^2/\mu_h^2  \, < \, b  \quad \mbox{whenever} \quad \alpha_h > 0$$
Then (\ref{RelBias}) holds approximately and takes the form
$$ E(\hat{V}^{\tt SDR})/V^{\tt tru} \, - \, 1 \; \approx \; \sum_{h=1}^H \,(N_h  \mu_h)^2 \,  (\alpha_h/n_h)^2 \, \Big/ \, \sum_{h=1}^H \, (N_h  \mu_h)^2 \, (\alpha_h/n_h) \, (\sigma_h^2/\mu_h^2+1-\alpha_h)$$
This setting is the only one we have found where RelBias would not become arbitrarily small with large domain sample size. It might occur if a domain occurred primarily in many isolated strata of varying sizes and the stratumwise domain proportions were roughly proportional to sample size, e.g., if $n_h$ is roughly proportional to the size $N_h$ of the population stratum.  However, even this example is not concerning as long as the average ratio ~$(\alpha_h/n_h)/(\sigma_h^2/\mu_h^2)$~ is no larger than $0.10$ to $0.15$.

In examples (ii)--(iii), extreme relative biases arise if the quantities $~q_h \, \equiv \, p_h \cdot \alpha_h \cdot \mu_h/n_h$ alternate in a long sequence of $H$ strata, but this requires a specific ordering of the strata. This ordering need only occur within the set of strata intersecting with a domain $A$ of interest. That might well happen for a few domains among many (e.g., if many small domains were of interest in a Small Area Estimation application) even if the ordering of strata were unexceptional. 

 Almost the only large relative biases in SDR variance estimates that we have found arise from examples where the ratios ~$q_h~$ alternate dramatically as in (ii)--(iii) above or where there are one or two strata with very large values of $q_h~$ and other values are much smaller. In the dramatically alternating examples, randomly permuting the order of strata generally restores the Relative Bias to be moderate even for small domains and to be roughly proportional to $1/n_A$. A preliminary step of permuting stratum indices to reduce SDR bias is worthwhile when the survey analyst has reason to believe that the quantities $q_h$  under a pre-existing ordering oscillate dramatically, and for that reason we include an {\tt R} function permuting strata within the small package of {\tt R} functions described in Appendix~\ref{rcode} to implement the Relative Bias formulas (for general $D$) introduced in this Section and proved in Appendices~\ref{prop.proof} and \ref{D.relBias}. On the other hand, Supplement~\ref{perm} shows through simulations of a variety of superpopulations and stratified SRS designs that replacing the SDR variance estimate with the median of the SDR estimates from a large number of random permutations of stratum order (only when VSDR exceeds the median is not generally an effective way to improve the accuracy of SDR-based variance estimation.

\subsection{Simulations Showing Typical SDR Relative Biases} \label{sims}

Next, we study the behavior of relative bias in SDR variance estimation across population and stratified sample-design and domain parameters by factorially generating a variety of superpopulation and stratified SRS scenarios. The goal is to exhibit and confirm the typical behavior of formula (\ref{VSDR2})  and to explore via simulation the SDR bias under different numbers $D$ of cycles. The method involves generating sets of parameters $(\mu_h, \alpha_h, \; h=1,\ldots,10)$ from probability distributions, using Gaussian copulas with specified correlations to induce (exchangeable) dependence across strata.

In the factorially generated scenarios, $H=10$ and for all $h=1,\ldots, 10$, \ \  $N_h= N/H = 10^5$ and $n_h=n/H$ are balanced across strata, and $~\sigma_h^2 = 1$. Mean-parameters $ (\mu_h, \; h=1,\ldots, 10)$ were all generated randomly as vectors with ~Gamma$(3, \, {\tt rate=}0.5)$ marginals and dependence according to a Gaussian equicorrelated copula with correlation $\rho_{\mu} = 0.3$. Expected domain parameters $\alpha_h^{\circ}$ were generated marginally as ~Beta$(a,b)$ with 6 combinations of $(a,b)$ parameters such that the mean is $a/(a+b) = $\linebreak

\renewcommand{\baselinestretch}{1.1}
\begin{figure}[h]
    \begin{center} \vspace{3mm}
        \includegraphics[width=5.4in]{"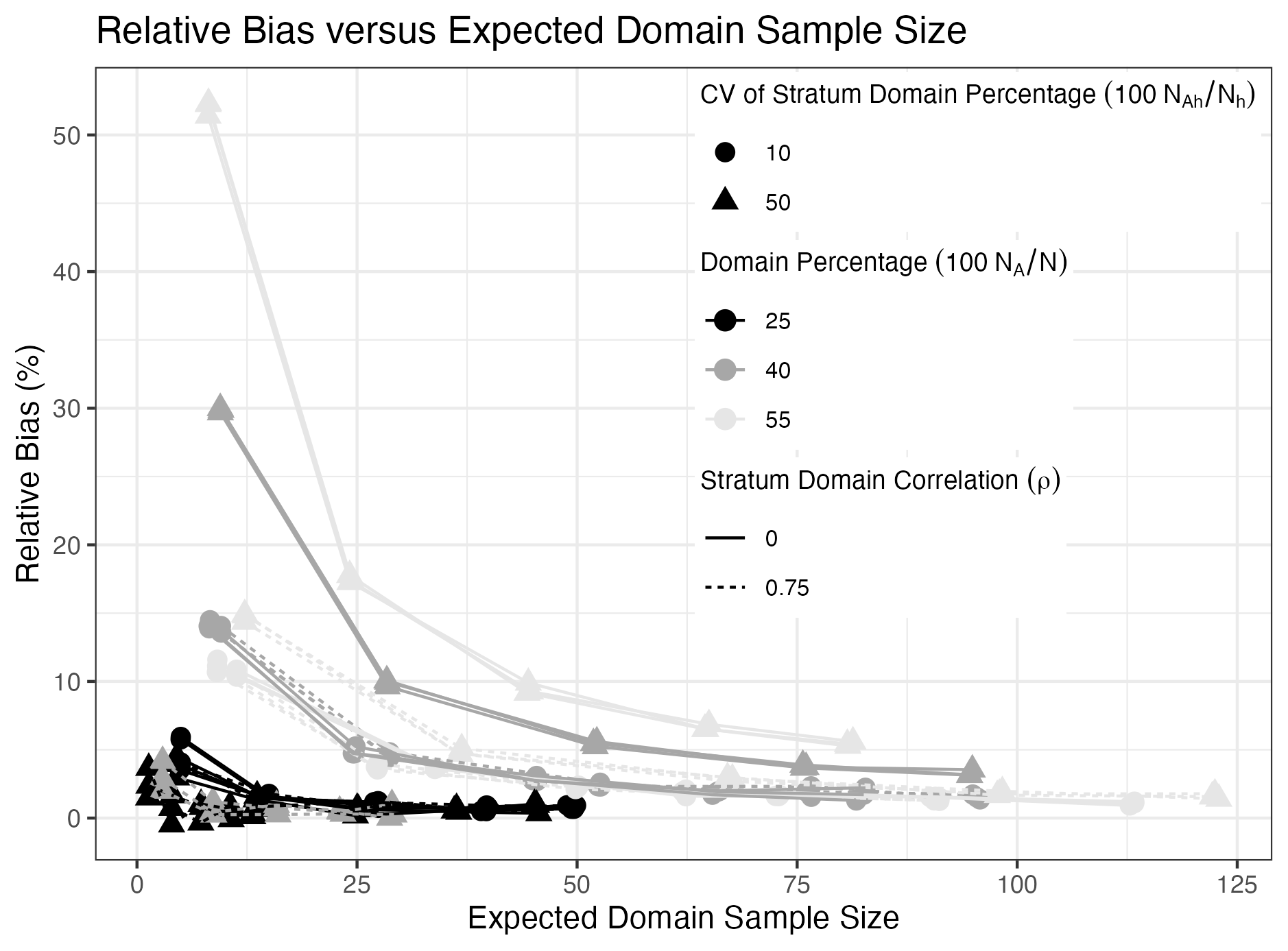"}
    \caption{Average relative biases, across 10,000 samples drawn by Stratified SRS at each of 5 sample sizes for each of 36 superpopulations. Superpopulations were generated with 3 independent replicates for 12 factorial parameter combinations, consisting of 3 levels of domain proportion $N_A/N$, \linebreak 2 levels of correlations (within a Gaussian copula)  and 2 levels of CV among stratum domain-proportions $\alpha_h^\circ$. Each superpopulation had stratum means generated as $\mu_h \sim \mbox{Gamma}(3, 0.5)$ correlated $0.3$ across strata, with {\tt iid\/} outcome variables $y_i \sim {\cal N}(\mu_h,1)$ in stratum $h$.} 
    \label{vanilla_plot}
    \end{center}
\end{figure} 
\renewcommand{\baselinestretch}{1.3} 

\noi $0.25, 0.40$, or $0.55$ and squared coefficient of variation is  $b/(a(a+b+1)) = 0.1$ or $0.9$, with dependence across strata specified by a Gaussian copula with correlation $\rho_{\alpha} = 0$ or $0.75$. For each such set of superpopulation parameters, 3 superpopulations were generated so that within stratum $h$, the domain indicators $\epsilon_i \, \sim \, \mbox{Bernoulli}(\alpha_h^{\circ})$ are {\tt iid}, independent of the {\tt iid} outcomes ~$y_i \, \sim \, {\cal N}(\mu_h, \sigma_h^2)$. For each of 5 different  sample sizes $n$, ~10,000 independent stratified SRS samples were drawn from each superpopulation.   Figure~\ref{vanilla_plot} shows a summary of these simulation results. Each curve shows 5 points for the 5 different sample sizes, with domain sample size and relative bias averaged across the $ 10^4$ samples drawn. The averaging across samples affects only the numerator of the relative bias, since the $V^{\tt tru}$ denominator is the same for each set of superpopulation parameters. 

 \begin{figure}[h]
     \begin{center} 
        \includegraphics[width=6.3in]{"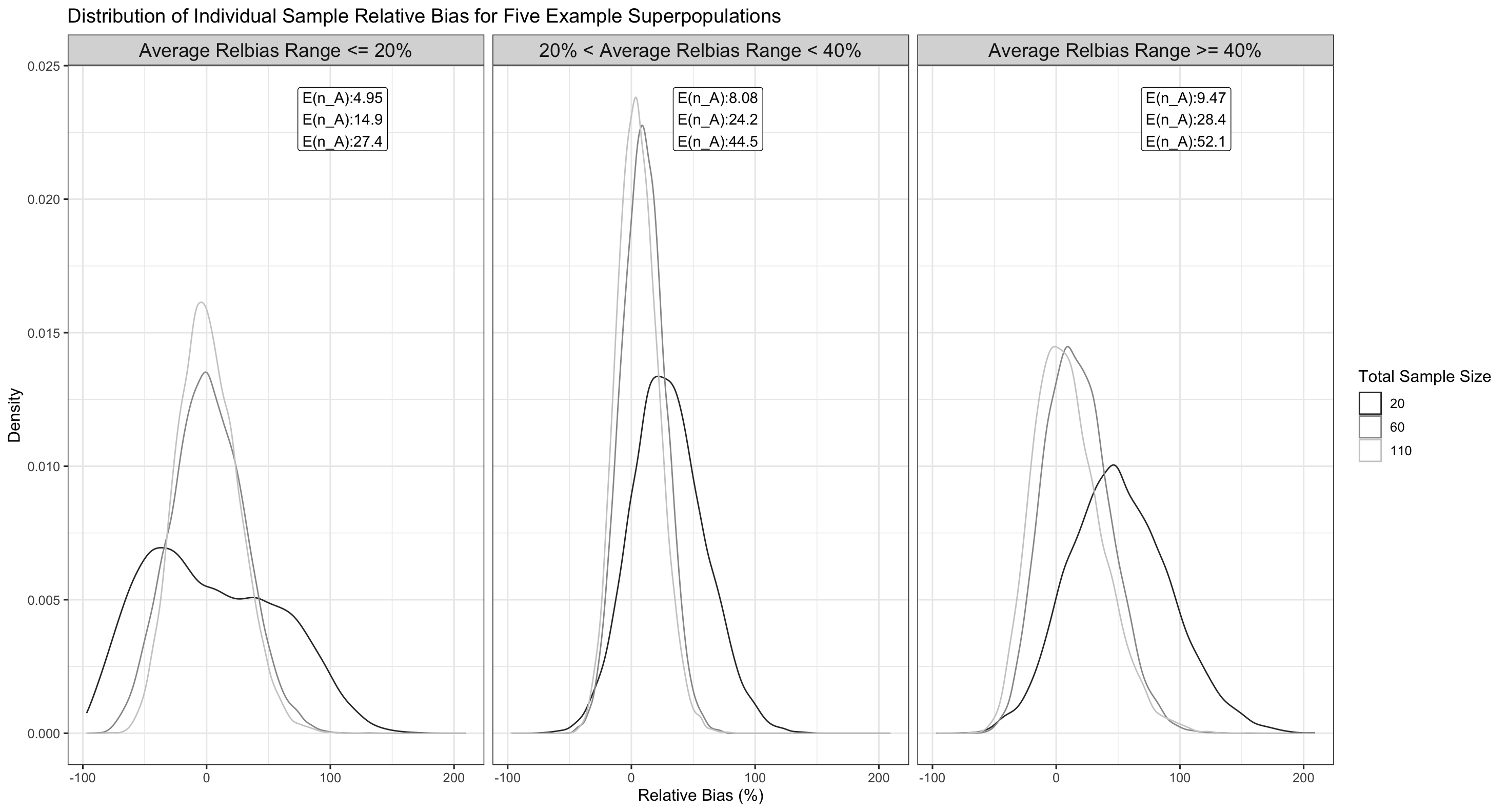"} \vspace{-3mm}
   \caption{Variance of SDR variance estimates across stratified SRS  samples drawn from $3$ of the  superpopulations defined in Figure~\ref{vanilla_plot}, at three different sample sizes $n$. The Fig.~1 superpopulations were stratified into three groups based on their average relative bias when $n=20$. One superpopulation was chosen at random from each category for this figure. Samples with $0$ domain elements were excluded from these density plots, with greatest impact on leftmost panel with    smallest expected domain sample. SDR variance estimates are highly variable even when the expected relative bias is small. }
 \label{VariableRelBias} \vspace{-3mm}
  \end{center} \end{figure}

In Figure~\ref{vanilla_plot}, whenever the expected domain sample size is above 25, relative bias falls below 20\% even for the most extreme cases presented. Bias is largest when domain proportions $N_A/N$ are large, \linebreak 

\renewcommand{\baselinestretch}{1.1}
\begin{figure}[h]
    \begin{center}
        \includegraphics[width=5.6in]{"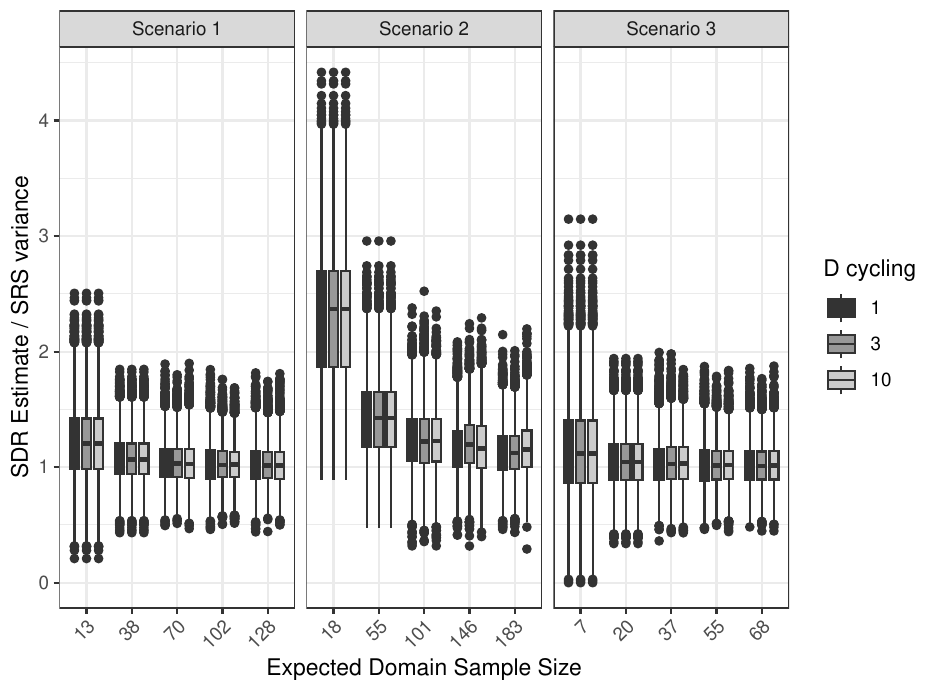"} \vspace{-2mm}
    \caption{Boxplots for  ratios $\hat{V}^{\tt SDR}/V^{\tt tru}$ of SDR variance estimates over true stratified SRS variances for 10,000 stratified SRS samples drawn from each of 3 superpopulations ({\it Scenarios\/}) generated randomly as in Sec.~\ref{sims}, each with 5 distinct expected sample sizes as indicated. For each superpopulation and sample, estimates were calculated three times using $D$-cycling parameters $1,3, 10$. Boxplots are displayed for common expected domain sample size for comparison across $D$. Results across $D$ are strikingly equivalent  for expected domain sample size $\le 75$, and still very much the same across $D$ despite some variability for larger domain sample sizes.}
    \label{Dcycling_plot}
    \end{center}
\end{figure} \vspace{-3mm}
\renewcommand{\baselinestretch}{1.3}

\noi as suggested in Example (iv),  and the stratum domain correlation is $0$ which we may interpret as inducing the most pronounced alternation (as in Example (ii)) among parameters $q_h$). The relative bias is also affected by the overall coefficient of variation of $\alpha_h^\circ$, with larger CVs allowing for greater $q_h$-alternation. These effects diminish as the expected domain sample size increases. However,
while these visual patterns are typical for most choices of parameters within the stratified SRS sampling of {\tt iid\/} superpopulations, the expected SDR variance estimates reflect serious relative biases for some parameter combinations. 


SDR variance estimates for individual samples are highly variable, as exhibited in Figure~\ref{VariableRelBias}. To create that Figure, the set of superpopulation parameter sets used in constructing Fig.~\ref{vanilla_plot} was subdivided into 3 categories according to ranges of average relative bias associated with a stratified SRS overall sample size $n=20$. One superpopulation parameter set was selected from each category and a single superpopulation generated from those parameters. Fig.~\ref{VariableRelBias} shows density estimates for the ratio of these single-sample SDR variance estimates divided by {\tt Vtru} minus $1$, labeling them as `relative bias'. Similar pictures generated with many samples drawn from other superpopulations show a similar phenomenon. For small to moderate sample sizes, the range of SDR estimates relative to {\tt Vtru} is really broad.

\subsection{D-Cycling}  \label{Dcycling}

We now turn to the effect of the D cycling parameter, using superpopulations and samples generated as in  Section~\ref{sims}. For each sample, SDR estimates were calculated for each of 3 choices of D cycling parameter, $D=1,3$, or $10$, using the permutation sequences $\mathbf{a}^{(d)}$ given in Appendix~\ref{cycle10}. Figure~\ref{Dcycling_plot} shows \noi the distribution of SDR estimates batched by expected domain sample size, pooled across samples for three superpopulations (`scenarios'). These pictures are typical of those generated for a variety of other single superpopulations. (
Supplement~\ref{morepops} provides additional examples, with the same implication.)  
The biases arising from SDR with these different $D$ values look remarkably the same within the triplets of boxplots within each bin of expected domain sample size, and this is our experience across a range of expected domain sample sizes and many different scenarios. The range exhibited by the boxplots indicates the variability of the SDR variance estimates, shown here as ratios over the true stratified SRS variances $V^{\tt tru}$. (Recall that the SDR estimates are equivalent by definition for all $D$ when $n \le 78$, and this is true for some superpopulations in which the $n_A$ values are small.)

\section{Discussion} \label{Disc}

Our formulas and simulations do not address clustered structure in the underlying finite population and samples, nor did any previous published studies of SDR bias. The stratified-SRS design is assumed in order to obtain tractable formulas, but it plays the dual role of expressing how sort-ordering the respondent population can affect the SDR variance and how (through $q_h$ parameters) the differences between successive segments of the population should be quantified. In surveys that are done repeatedly, it will be known for some outcome variables (those that naturally cluster in address neighborhoods) that $q_h$ values are fairly stable across neighborhoods, and SDR variance estimation should work well. For other outcome variables, results from previous surveys may suggest a sort-ordering of samples that is better than the default address-based ordering for SDR-estimation. However, the fact that the essential parameters $\mu_h, \, \sigma^2_h$ are features of the outcome-attribute while $\alpha_h$ is a feature of the domain of interest makes it hard to recommend a parameter-based strategy for reordering of respondents even if the parameter values can be guessed approximately.

The theory developed here could be applied to surveys with {\tt  iid\/} clusters as units, thereby augmenting the justification for SDR variance estimation in real surveys. This is actually done in CPS (U.S.~Census Bureau,  2019, pp.~53, 93) where replicate weight factors are defined at the level of 4-household clusters defined as respondents when at least one of the 4 houserholds responds. Replicate weighting with clusters defined as units has similarly been used to justify bootstrap-based  variance estimation (Rao and Wu 1988).

The distinction among the SDR variance estimates produced with different cycling numbers $D$ make some degree of $D$-cycling (with $D>1$) worthwhile. Section~\ref{SDR.Variance} showed a slight but definite advantage for $D \ge 3$ in terms of variance of SDR variance-estimates, and the bias of the SDR variance estimates is no worse with $D \ge 3$ than with $D=1$.

\subsection{Relevance of SDR to SAE Modeling} \label{SAE}

SDR is the standard methodology for variance estimation in large household surveys (ACS, CPS) conducted by the US Census Bureau. That also makes it the standard method for deciding when domains are too small to provide reliable survey-weighted estimates  of population attribute totals. In some settings where estimates are needed in smaller domains, the Census Bureau publishes estimates using model-based Small Area Estimation methods, and these methods use small-domain SDR variance estimates as inputs  (Rao and Molina 2015). In many Small Area Estimation problems, `linking models' connecting the different small areas are based on exact or approximate knowledge of the variances of small-area survey-weighted totals (Rao and Molina 2015). Huang and Bell (2009) stated that in Census Bureau applications of Small Area Estimation using ACS data, small-domain variances are generally derived from replicate-based estimates and then often assumed known possibly after replacing the variances by values fitted to a Generalized Variance Function or GVF (Wolter 2007). Further description of GVFs as used at the US Census Bureau in the Small Area Income and Poverty Estimates (SAIPE) program can be found in Maples et al.~(2009) and Bell et al.~(2016). One of the few references to study the sensitivity of SAE estimates to biases in estimated variances is Bell (2008), finding that the estimates are seriously affected only when biases are large and the focus is on conditional predictions in individual small areas. However, this is true in important statistical agency applications of SAE.

An important Small Area Estimation program at the Census Bureau is the estimation from ACS and decennial-census data of totals and ratios in small populations based on geography, race and language proficiency supporting the determinations of statutorily mandated ballot assistance under the Voting Rights Act (Census Bureau 2018, 2022). Variance estimation in the small-area model-based estimates in that program served as the specific motivation behind this research (Slud and Ashmead 2017). The need for a general study like this one arose because the behavior of SDR methods in variance estimation (for totals) in small domains has not been studied previously, beyond investigations of Huang and Bell (2009, 2010) with respect to equally weighted {\tt iid} superpopulations and the simplest case of SDR with fixed $R$ and without multiple cycles.

The statistical agency surveys used for SAE are generally large, with larger geographic areas (regions, sometimes states or large counties or metropolitan sub-state areas) containing sufficient sample that there would be no harm in thinking of them as strata or post-strata. The interesting domains in SAE often cut across many such strata, and that is the mental model for our study of SDR in stratified surveys, although the statistical-agency surveys never sample by SRS within those strata. Our initial premise, following Huang and Bell (2009), was that the bias of SDR estimates of variance should generally be small for surveys and domains with domain proportions $\alpha_h$ mostly very small. However, many SAE studies of interest to statistical agencies are based on many domains (based on demographic cells or economic characteristics) of moderate size (up to $100$, say, as in VRA), and our study -- particularly the special examples (ii), (iii) and maybe (v) -- suggest that a few of those domains may have biases in SDR variance estimates in the range 20--50\%, and in any case the small-domain variance estimators themselves are highly variable. That may not be concerning for most small-area estimates, but it will lead to serious prediction errors for some of them. 

Within CPS, Trudell et al.~(2018) noted an alternating pattern in SDR variance estimates induced by the lexicographic sort of respondents by demographic block characteristics
which did not respect the time-ordered nature of CPS sample. This anomaly arose in part because CPS weight-replicates are defined in terms of baseweights, at the sampling stage (U.S.~Census Bureau, 2019, p.~82), rather than on the file of respondents as this paper assumes.  Correcting this led to a decrease in variance estimates within Self-Representing PSUs. More broadly, domain patterns that lead to isolated 'strata' due to the sorting of blocks by area level characteristics may induce SDR bias, but this requires further study.

The noisiness of variance estimates in small domains is not remediable, but it may be that the geographical (generally address-based) ordering of sampled units leads to alternating or isolated subsets of domains (in the sense of examples (ii) or (v)) and thereby to unnecessarily large upward biases in variance estimates. Automatic tools based on the permutation of address blocks would be relatively easy to embed in the code for estimating variances, at least for statistical agency SAE studies. However, as Supplement~\ref{perm} shows, that approach can introduce new biases, and we do not recommend it.

\subsection{Recommendations} \label{recs}

Considerations of variance and bias of SDR variance-estimators lead us to recommend $D=3$ or $4$, along with $R=80$ as the largest number of replicates that can generally be retained and published in large surveys based on costs of computation and storage. Some large surveys, like CPS, do retain and publish 160 replicates in (sampled subsets of) microdata, and R as large as that can be beneficial for bias and variance. With any choice of $D$ and $R$, SDR tends to overestimate variances in small domains, an effect that should be accounted for in SAE applications using ACS and CPS.

It may be advisable to order geographic units (in lieu of `strata')  to respect the ordering of $q_h$ estimates from previous survey runs, when that is possible. However, if the strata reflect address ordering in large statistical-agency surveys, then we should be wary of permuting them. Our numerical experiments (compare Supplement~\ref{perm}) suggest that this is not helpful, i.e., can introduce negative bias, to reduce SDR variance estimates when they are larger than the median or another quantile of the set of SDR variances obtained from many random permutations of strata.

For domain estimates in SAE based on larger surveys: if a domain is such that many strata do not contain any domain elements, it may be sensible to treat those $0$'s as structural, i.e., to omit sampled units from those strata in calculating variance using SDR. That will reduce the bias of SDR estimates. Moreover, for some outcome variables it may be natural to treat the sample in a domain as the whole sample for purpose of SDR-based variance estimation. Although that is \emph{not} currently done for small-domain estimates in agency SAE projects like VRA, it easily could be and would tend to reduce upward bias  of variance estimates.

\newpage
\section*{References}
\begin{description}
\item American Community Survey (2022), Survey Design and Methodology document, version 3.0, Ch.~12 Variance Estimation.
\url{https://www2.census.gov/programs-surveys/acs/methodology/design_and_methodology/2022/acs_design_methodology_ch12_2022.pdf} 

\item Ash, S. (2014), Using successive difference replication for estimating variances, {\it Survey Methodology\/} {\bf 40}, 47-60. 

\item Bell, W.~(2008), Examining Sensitivity of Small Area Inferences to Uncertainty About Sampling
Error Variances, {\it Proc. Amer.~Statist.~Assoc., Survey Res.~Meth.~Section\/}, 327-334.

\item Bell, W., Basel, W. and Maples, J.~(2016), An Overview of the U.S. Census Bureau's Small Area 
Income and Poverty Estimates Program, Chap.~19 in: {\bf Analysis of Poverty Data by Small Area 
Estimation}, ed.~M.~Pratesi, John Wiley.

\item Fay, R. (1984), Some properties of estimates of variance based on replication methods, 
{\it Proc. Amer. Statist.~Assoc., Survey Res.~Meth.~Section\/}, 495-500.

\item Fay, R. and Train, G. (1995), Aspects of survey and model-based post-censal 
estimation of income and poverty characteristics for states and counties, 
{\it Proc.~Amer.~Statist.~Assoc., Govt.~Statist.~Section\/}, 154-159. 
\url{https://cps.ipums.org/cps/resources/repwt/FayTrain95.pdf}

\item Huang, E. and Bell, W. (2009) A simulation study of the distribution of Fay's Successive \linebreak Difference 
Replication variance estimator, {\it Proc.~Amer.~Statist.~Assoc., Survey Res.~Meth. Section\/}, 212-217.

\item Huang, E. and Bell, W. (2010) Further simulation results on the distribution of some survey variance estimators, 
{\it Proc.~Amer.~Statist.~Assoc., Survey Res.~Meth.~Section\/}, 3877-3889.

\item Krewski, D. and Rao, J. (1981), Inference from stratified samples: properties 
of the linearization, jackknife and balanced repeated replication methods, {\it 
Ann.~Statist.\/} {\bf 9}, 1010-1019.

\item Maples, J., Bell, W. and Huang, E. (2009), Small area variance modeling with application to county
poverty estimates from the American Community Survey, {\it Proc.~Amer.~Statist. Assoc., Survey 
Res.~Methods Section\/}, 5056-5067.

\item Navarro, A. (2001), 2000 American Community Survey (ACS) Comparison County Replicate Factors,
internal Memorandum (ACS-V-01) to C. Alexander, US Census Bureau, prepared by K. Albright, May 23, 2001.

\item Plackett, R.~and Burman, J.~(1946), The design of optimum multifactorial experiments, {\it Biometrika\/} 
{\bf 33},  305-325 .

\item Opsomer, J., Breidt, J., White, M. and Li, Y. (2016), Successive Difference Replication variance 
estimation in two-phase sampling, {\it Jour.~Survey Statist.~\& Methodol.\/} {\bf 4}, 43-70.

\item  R Core Team (2023), {\bf R: A language and environment for statistical computing}. \linebreak R
  Foundation for Statistical Computing, Vienna.  \url{http://www.R-project.org/}.

\item Rao, J. and Molina, I. (2015) {\bf Small Area Estimation}, 2nd ed., Wiley.

\item Rao, J.  and Wu, C.-F.(1988). Resampling inference with complex survey data. {\it Jour,~Amer.~Statist.
Assoc.\/} {\bf 83}, 231-241.

\item Slud, E., Ashmead, R., Joyce, P.~and Wright, T.~(2018), Statistical modeling methodology (2016) for Voting
Rights Act, Section 203 determinations. Technical report, US Census Bureau. \url{https://www.census.gov/content/dam/Census/library/working-papers/2018/adrm/RRS2018-12.pdf}

\item Slud, E. and Ashmead, R. (2017), Hybrid BRR and Parametric-Bootstrap Variance Estimates for 
Small Domains in Large Surveys, {\it Proc.~Amer.~Statist.~Assoc., Survey Res.~Methods Section\/}, 1716-1730.

\item Slud, E., Franco, C., Hall, A.~and Kang, J.~(2022), Statistical modeling methodology (2021) for Voting
Rights Act, Section 203 determinations. Technical report, US Census Bureau. \url{https://www.census.gov/content/dam/Census/library/working-papers/2022/adrm/RRS2022-06.pdf}

\item Slud, E., Franco, F.~and Hall, A.~(2024), Small area estimates for Voting Rights Act Section 203(b) coverage 
determinations, {\it Calcutta Statistical Association Bulletin\/} {\bf 76}(1), 137--159.

\item Sukasih, A. and Jang, D. (2003) Monte Carlo study on the Successive Difference Replication 
method for nonlinear statistics, {\it Proc.~Amer.~Statist.~Assoc., Survey Res.~Methods 
Section\/}, 4141-4147.


\item Trudell, T., Dong, K., Slud, E. and Cheng, Y. (2018), Computing Replicated Variance for Stratified 
Systematic Sampling,  Joint Statistical Meetings talk, Vancouver, CA. 
 
\item   U.S. Census Bureau,  Current Population Survey  Design and Methodology,  Technical Paper 77,  October 2019.

\item Wolter, K. (1984), An investigation of some estimators of variance for systematic sampling, {\it 
Jour. Amer.~Statist.~Assoc.\/} {\bf 79},  781-790.

\item Wolter, K. (2007) {\bf Introduction to Variance Estimation}, 2nd ed., Springer.

\end{description}

\section*{\Large Appendices}
\appendix
\section{Implementations of 10-cycle Indexing in the ACS} \label{cycle10}

The indexing (\ref{fdefD}) implemented in ACS has parameters $R=80, \, m=78, \, D=10$. Navarro (2002), 
Sukasih and Jang (2003), and Ash (2014) all describe in words a similar idea for constructing $10$ cycles $\mathbf{a}^{(d)}$ of row-indices from the row-numbers $\{1,\ldots, 80\}$ of an $80\times 80$ Hadamard matrix (omitting rows $1, 41$). That idea is to build the $d$'th cycle from arithmetic progressions of span $d$, re-cycling as necessary to use all indices  in ${\cal A} = \{2,\ldots, 40, 42,\ldots, 80\}$. Simple {\tt R} code to do this is as follows: \vspace{-2mm}
\begin{verbatim}PermArray = array(0, c(78,10))
PermArray[,1] = c(2:40,42:80)
for(j in 2:10) {
       cyc = NULL
       for(k in 1:j) cyc = c(cyc, seq(1+k,80,by=j))
       PermArray[,j] = setdiff(cyc,41) }
\end{verbatim}
The resulting columns $\mathbf{a}^{(d)}, d=1,\ldots, 10$ are easily generated and are given explicitly in Supplement~S1. This may have been the set of 10 index blocks intended by Sukasih and Jang (2003); 
they did not specify precisely. Among the 780 row-index pairs generated by (\ref{fdefD}) 
using these cycles, it turns out that there are exactly 754 unique pairs, so that 26 are duplicates.

The set of 10 cycles of 78 actually used in ACS indexing for SDR is generated using the SAS code by 
K.~Albright given in Navarro (2002). Running that code and renumbering rows to run from 2 to 80 
(using the rule  $k \mapsto 1 +( (k-1) \mod 80$) yields the 10 cycles explicitly listed as $\mathbf{a}^{*(d)}, \; d=1,\ldots, 10,$ in Supplement~S1. These 10 index cycles $\mathbf{a}^{*(d)}$, result in 780 row pairs of which 
758 are unique and 22 are duplicates. Of the pairs that occur more than once: (80,3), (39,2), (40,43) 
and (79,42) each occur four times; and (38,4), (78,44), (35,4), (36,5), (37,6), (38,7), (75,44), 
(76,45), (77,46), and (78,47) occur twice. The set of 7 cycles  $\mathbf{a}^{*(d)}$ for $d=1,2,4,5,7,8,10$ 
would produce a set of $7 \cdot 78 = 546$ unique row-pairs with no duplicates at all, but that including 
cycles $\mathbf{a}^{*(d)}$ for $d=3,6$ or $9$ would produce 8 cycles that do induce row-pair duplicates.

\section{$M, Q$ and the Variance of $\hat{V}^{\tt SDR}$}   \label{App2.Var}

\subsection{Sparse Representation of the Matrix $M$}

Section~\ref{SDR.Variance} defined the $(j,j')$ entry of the $n \times n$ matrix $M$ under cycling-number $D$ with cycling permutations $(\gamma_j, \; j=1,\ldots, Dm+1)$ by the right-hand side of equation (\ref{Qalt.D}). (Recall the preliminary modification that $\gamma_{n+1} \equiv \gamma_1 = a_1$ when $D=1$ and $n<m$.) Generally, most entries of $M$ are $0$, so we compute quantities related to $\hat{V}^{\tt SDR}$ and its mean and variance through a sparse-matrix representation $\tilde{M}$ (computed by the function {\tt Mcalc} in the package described in Appendix~\ref{rcode}) containing all rows 
$$\big(~j, \; j', \;  \frac{1}{2}\, \big[ I_{[\gamma_j = \gamma_{j'}]} \, + \, I_{[\gamma_{j+1} = \gamma_{j'+1}]}  \, - \,
I_{[\gamma_j = \gamma_{j'+1}]} \, -  \, I_{[ \gamma_{j+1} = \gamma_{j'}]} \big] \big)$$
for which the third entry is non-zero, with rows ordered lexicographically with respect to pairs $(j,j') \in \{1,\ldots, n\}^2$. The matrices $M, \, \tilde{M}$ are combinatorial objects, depending only on $m, n, D, \{\gamma_j\}_{j=1}^{n+1}$.

The rows of $\tilde{M}$ with respective first two entries $(j,j')$ and $(k,k')$ have identical third entries whenever $(\gamma_j, \gamma_{j+1}) = (\gamma_k, \gamma_{k+1})$ and $(\gamma_{j'}, \gamma_{j'+1}) = (\gamma_{k'}, \gamma_{k'+1})$, and in that case, with $g$ representing the equivalence class of $j$ and $g'$ representing the equivalence class of $j'$, formula (\ref{Qalt.D}) defines $Q_{g,g'} = M_{j,j'} = M_{k,k'}$.

Formula (\ref{VsdrG}) defining $\hat{V}^{\tt SDR} \, = \, \hat{V}^{\tt SDR}(\hat{Y})$ is re-expressed directly without equivalence classes $g,g'$ by 
\beq \hat{V}^{\tt SDR}(\hat{Y}) \, = \, \sum_{j=1}^n \, \sum_{j'=1}^n \, M_{j,j'} \, \breve{y}_j^\circ \,  \breve{y}_{j'}^\circ 
\label{VSDRhat.2} \eeq
The latter sum over $(j,j')$ gives the same answer when restricted to the rows $t$ of $\tilde{M}$, with $j \equiv j(t) =\tilde{M}_{t,1}, \;  j'\equiv j'(t) =\tilde{M}_{t,2}$, and $M_{j,j'} = \tilde{M}_{t,3}$.

\subsection{Variance of $\hat{V}^{\tt SDR}$ for $y_i$ {\tt iid\/} under SRS sampling}

Consider now the calculation of $var(\hat{V}^{\tt SDR})$ when the superpopulation outcomes $y_i$ are  {\tt iid\/} and sampling is SRS. This variance depends only on the moments $E\{ y_1^r \}$ for $r=1, \ldots, 4$ and the matrix $M$ or $\tilde{M}$. Let $m_r =  E(y_1^r)$ and $\mu_r = E\big\{ \big(y_1 \, - \, m_1 \big)^r\}$ for $r \le 4$ with $\mu = m_1, \, \sigma^2 = \mu_2$, and let $\ul{\bf 1} = \ul{\bf 1}_n \in \mathbb{R}^n$ denote the vector with all entries $1$. 
Then the nonzero contributions of 
$$ var\big\{ (\hat{V}^{\tt SDR})^2 \big\} \, = \, \sum_{j=1}^n \,  \sum_{j'=1}^n \,  \sum_{k=1}^n \,  \sum_{k'=1}^n \, M_{j,j'}  \, M_{k,k'} \, \big\{ E\big[ \breve{y}_j^\circ \, \breve{y}_{j'}^\circ \, \breve{y}_k^\circ \, \breve{y}_{k'}^\circ \big] \, - \, \frac{N^4}{n^4} \, (\mu^2 + \delta_{j,j'} \sigma^2) \,  (\mu^2 + \delta_{k,k'} \sigma^2)\big\}$$
can be decomposed as the sum of four terms 
$$= \, \frac{N^4}{n^4} \, \big[T_1 \cdot (m_4 - (\mu+\sigma^2)^2) \, + \ , T_2 \cdot (m_3 \mu - \mu^2(\mu^2+\sigma^2) \, + \, T_3\cdot (m_2^2 - \mu^4) \, + \, T_4\cdot(m_2\mu^2 - \mu^4)\big]$$ 
where, using summetry of $M$ and $M_{j,j} = 1$ for all $j$, and with $|A|$ denoting  the number of distinct elements  of a set $A$ of indices,
\begin{eqnarray*} T_1 & = & \sum_{j,j',k,k'}\,  M_{j,j'} \, M_{k,k'} \, I_{[ j=j'=k=k']} \; = \; n \\
T_2 & = & \sum_{j,j',k,k'} \, M_{j,j'} \, M_{k,k'} \, I_{[3 \mbox{\ of \ }  j, j',k,k' \mbox{\ identical \ }]} \; = \; 4 \, \ul{\bf 1}^{\tt tr} M \ul{\bf 1} \\ 
T_3 & = & \sum_{j,j'} \, M_{j,j'} \, (M_{j,j'}+M_{j',j}) \, I_{[j\ne j']} \; = \; 2 \, \big[\sum_{j,j'} \, M_{j,j'}^2 \, - \, n\big] \\
T_4 & = & \sum_{j,j',k,k'} \, M_{j,j'} \, M_{k,k'} \, I_{[ \, |\{j,j',k,k'\}| = 3]} \; = \;  4 \,  \sum_{j,j',k \mbox{ \ distinct}} \,  M_{j,j'} \, M_{k,k'}
\end{eqnarray*} 
 The computation of the terms $T_1, T_2, T_3, T_4$ as a function of $m, n, D$, and $\{\gamma_j\}_{j=1}^{n+1}$ is implemented in the {\tt R} function  {\tt VarSDRVar} in the {\tt R} package {\tt SDR} described in Appendix~\ref{rcode}. Then, in terms of those combinatorial coefficients, the variance of the $\hat{V}^{\tt SDR}(\hat{Y})$ estimator is given by 
\beq \frac{n^4}{N^4} \,  var\big(\hat{V}^{\tt SDR} \big) \, = \, \mu_4 \, T_1 \, + \, (\mu \mu_3) \, (4T_1+T_2) \, + \, (\mu^2\sigma^2) (4T_1+2T_2+2T_3+T_4) \, + \, \sigma^4 \, (T_3-T_1)  \label{VSDRvar} \eeq
We calculated this quantity for various combinations of $m, D$, and $n$. An important set of special cases, associated with Table~\ref{vartab1}, is where $K = n/m$ is an integer, and in those cases it turns out that for each $K = 1,\ldots, 10$, the only term of (\ref{VSDRvar}) that varies with $D$ is $T_1$, the coefficient of $\mu_4$, and the coefficients of $\mu \, \mu_3$ and $\mu^2\sigma^2$ are $0$. Moreover, by definition of the weight-replicates under cycling with parameter $D$, for fixed $n$ and $m$ the variance $var\big\{ (\hat{V}^{\tt SDR})^2 \big\}$ is identically the same for all $D \ge n/m$. Section~\ref{SDR.Variance} provides further calculations when  $\mu_4/\sigma^4 = 3$ (as is true for the normal distribution) showing that the variance of $\hat{V}^{\tt SDR}$ does not vary much for non-integer $n/m$ as long as $D \ge 3$.

A final technical remark is that for $D=1$, formula (\ref{VsdrG}) implies that $\hat{V}^{\tt SDR}/N$ is exactly unbiased for $var(\hat{Y}/N) = \sigma^2/n$ when (the sampling fraction $n/N$ is negligible and) $n \le m$ or $n =Km$ for integer $K$, but for $n>m$ not an integer multiple of $m$, the bias of $\hat{V}^{\tt SDR}/N$ is $\mu^2/n^2$.

\section{Proof of Proposition~\ref{Vform}} \label{prop.proof}

When $n \le m$, formula (\ref{VSDR2}) simplifies (with $g=j, \, Y^{(g)} = \breve{z}_j^{\circ}, \, G=n$) as in 
formula (\ref{SDGp})  to 
$$ N^{-2} \, \hat{V}^{\tt SDR} \; = \; \frac{1}{2N^2} \, \sum_{j=1}^n \, (\breve{z}_j^{\circ} - 
\breve{z}_{j+1}^{\circ})^2$$
with $\breve{z}_{n+1}^{\circ} \equiv \breve{z}_1^{\circ}$ and $h^{\circ}(n+1) \equiv h^{\circ}(1)$ by 
convention, and $z_j^{\circ} = \epsilon_j^{\circ} \, y_j^{\circ}$.

Properties {\bf (P.i)}--{\bf (P.iii)} provide first and second moments of the sample-indexed attributes and domain indicators  $\epsilon_j^{\circ}, \, y_j^{\circ}$.  These properties apply to give expectations in the last formula, in the form
\beq N^{-2} \, E\big(\hat{V}^{\tt SDR}\big) \; = \; \frac{1}{2N^2} \, \sum_{j=1}^n \, E \Big\{ (\breve{y}_j^{\circ} - 
\breve{y}_{j+1}^{\circ})^2 \, \cdot \, \epsilon_j^{\circ} \epsilon_{j+1}^{\circ} \, + \, \breve{y}_j^{\circ 2} \, \epsilon_j^{\circ}(1- \epsilon_{j+1}^{\circ}) \, + \, \breve{y}_{j+1}^{\circ 2} \, \epsilon_{j+1}^{\circ}(1- \epsilon_j^{\circ})\Big\} \label{EVsdr0} \eeq
It follows immediately from {\bf (P.i)} that 
$$E(\breve{y}_j^{\circ} - \breve{y}_{j+1}^{\circ})^2  \, | \, \epsilon_j^{\circ}=\epsilon_{j+1}^{\circ}=1) \, = \, 
 (w_j^{\circ} \, \mu_{h^{\circ}(j)} - w_{j+1}^{\circ} \, \mu_{h^{\circ}(j+1)})^2 \, + \,  (w_j^{\circ} \, \sigma_{h^{\circ}(j)})^2 \, + \,  (w_{j+1}^{\circ} \, \sigma_{h^{\circ}(j+1)})^2 $$
$$E(\breve{y}_j^{\circ 2} \, | \, \epsilon_j^{\circ}=1, \epsilon_{j+1}^{\circ}=0) \, = \, w_j^{\circ 2} \, (\mu_{h^{\circ}(j)}^2 + \sigma_{h^{\circ}(j)}^2)  $$
$$E(\breve{y}_{j+1}^{\circ 2} \, | \, \epsilon_{j+1}^{\circ}=1, \epsilon_j^{\circ}=0) \, = \, w_{j+1}^{\circ 2} \, (\mu_{h^{\circ}(j+1)}^2 + \sigma_{h^{\circ}(j+1)}^2)  $$

When $D=1$, sequential ordering of sample elements and fixed sample size $n_h$ within stratum $h$ implies 
that for each $h=1,\ldots, H$, there are $n_h-1$ pairs $(j,j+1)$ with $h^{\circ}(j) = h^{\circ}(j+1) = h$ and one pair with $h^{\circ}(j) = h \ne h^{\circ}(j+1) = h+1$. We account for these different types of terms in (\ref{EVsdr0}), using {\bf (P.ii)} to compute probabilities for $\epsilon$ product terms within a single stratum $h$, and using {\bf (P.iii)} for $\epsilon$  product terms across two strata. The result on the right-hand side of (\ref{EVsdr0}) is  
$$  \frac{1}{2N^2} \; \sum_{h=1}^H \; \Big\{ (n_h-1) \, \frac{N_h^2}{n_h^2} \, \Big[ 2 \alpha_h \, \frac{N_{Ah}-1}{N_h-1} \, \sigma_h^2 \, + \, 2 \, \alpha_h(1-\alpha_h) \frac{N_h}{N_h-1} \, (\mu_h^2+\sigma_h^2) \, \Big] $$
$$ \qquad + \; \alpha_h \, \alpha_{h+1} \; \Big[ \Big(\frac{\mu_h \, N_h}{n_h} \, - \, \frac{\mu_{h+1} \, N_{h+1}}{n_{h+1}}\Big)^2 \, + \, \frac{N_h^2 \, \sigma_h^2}{n_h^2}  \, + \, \frac{N_{h+1}^2 \, \sigma_{h+1}^2}{n_{h+1}^2}\Big] $$
$$ \qquad \qquad + \; \alpha_h \, (1-\alpha_{h+1}) \, \frac{N_h^2}{n_h^2} \, (\mu_h^2+\sigma_h^2) \; + \; 
  \alpha_{h+1} \, (1-\alpha_h) \, \frac{N_{h+1}^2}{n_{h+1}^2} \, (\mu_{h+1}^2+\sigma_{h+1}^2)  \Big\} $$
First collect terms in the last formula with the ratios $(N_{Ah}-1)/(N_h-1)$ and $N_h/(N_h-1)$ respectively replaced by $\alpha_h$ and $1$, and then account for the replacement. The result is
\bea &&N^{-2} \, E\big(\hat{V}^{\tt SDR}\big) \; = \;  \frac{1}{N^2} \, \sum_{h=1}^H \, \Big[ \frac{N_h^2}{n_h} \, \alpha_h \, \sigma_h^2 \; + \; \frac{N_h^2}{n_h} \, \alpha_h \, (1-\alpha_h ) \, \mu_h^2 \; + \; \frac{N_h^2}{n_h^2} \, \alpha_h^2 \,  \mu_h^2  \label{EVsdr1} \\ && \quad - \, \alpha_h \alpha_{h+1} \, \mu_h \mu_{h+1} \, \frac{N_h \, N_{h+1}}{n_h \, n_{h+1}}\Big] \; + \; \frac{1}{N^2} \, \sum_{h=1}^H \, \frac{(n_h-1) \, N_h^2}{(N_h-1) \, n_h^2} \, \alpha_h(1-\alpha_h) \, \mu_h^2 \nonumber \eea
Further algebraic combination of terms, with substitution of $p_h=N_h/N$, yields \ $ N^{-2} \, E\big(\hat{V}^{\tt SDR}\big) \; = $
$$\sum_{h=1}^H \, \Big[ \frac{p_h^2}{n_h} \, \alpha_h \, \big( \sigma_h^2 \, + \, (1-\alpha_h) \, \mu_h^2\big) \, \big(1 \, - \; \frac{n_h}{N_h} \, + \, \frac{n_h}{N_h}\big) \; + \; \frac{1}{2} \, \big( \frac{p_h \, \alpha_h \, \mu_h}{n_h} \, - \, \frac{p_{h+1} \, \alpha_{h+1} \, \mu_{h+1}}{n_{h+1}}\big)^2$$
$$ \qquad \qquad + \sum_{h=1}^N \; \big(\frac{p_h}{n_h \, N}\big) \; \frac{(n_h-1) \, N_h}{(N_h-1) \, n_h} \, \alpha_h(1-\alpha_h) \, \mu_h^2$$
$$ = \; v_{\tt tru} \; + \; \frac{1}{2} \, \sum_{h=1}^H \, \big( \frac{p_h \, \alpha_h \, \mu_h}{n_h} \, - \, \frac{p_{h+1} \, \alpha_{h+1} \, \mu_{h+1}}{n_{h+1}}\big)^2 \, + \, \sum_{h=1}^H \, \frac{p_h}{N}\Big[\alpha_h \, \sigma_h^2 \, + \, \alpha_h (1-\alpha_h) \, \mu_h^2 \, \Big(1 \, + \; \frac{(n_h-1) \, N_h}{(N_h-1) \, n_h^2}\Big)\Big]$$
which is just the result of formula (\ref{VSDR2}). \hfill $\Box$

\section{Expected SDR Variance Formula } \label{D.relBias}

In this section, do not assume that $n \le Dm$ or that the pairs $(\gamma_j, \gamma_{j+1})$ are distinct for $j \le Dm$, but if $n > m$, then according to (\ref{fdefD}), ~$\gamma_{n+1} = \gamma_{(n\!\!\mod Dm) +1}$, while for $n \le m$, $\gamma_{n+1} = \gamma_1$. To keep notations simple, for $n \ge j > Dm$ regard the sequence $\gamma_j$ as being periodically continued, $\gamma_j \equiv \gamma_{1 + (j-1)\!\!\mod \!Dm}$.  In addition, define index-sets ${\cal C}_h = \{ c_{h-1}+1, \ldots, c_h\}$, ~ where $c_0 \equiv 0$ and for $h \ge 1, \;\; c_h \, = \, \sum_{k=1}^h \, n_k$. The sets ${\cal C}_h$ are the ordered indices $j$ for sampled units $y_j^\circ$ in stratum $h$.

The formula is now less convenient than (\ref{VSDR2}) because of the need to pre-process the sets of $(j,j')$ pairs satisfying the conditions in the various indicators on the right-hand side of (\ref{Vsdr3}).
 The expectation of formula (\ref{Vsdr3}) [or equivalently, of (\ref{VSDR.M}) or (\ref{VSDRhat.2})] using properties (P.i)--(P.iii) in Section~\ref{bias.theory} is given by
$$ E(\hat{V}^{\tt SDR}) \, = \, \sum_{h=1}^H \, \sum_{j=1}^n \, \sum_{j'=1}^n  \, \frac{N_h^2}{n_h^2} 
\, I_{[j,j' \in {\cal C}_h ]} \,  \Big\{ \,  I_{[j=j']} \, \alpha_h \, (\mu_h^2+\sigma_h^2) \, + \, I_{[j \ne j']} \,  \alpha_h \, \frac{\alpha_h \, N_h - 1}{N_h-1} \, \mu_h^2 \, M_{j,j'} \Big\} $$
\beq  + \; \sum_{h=1}^H \, \sum_{h'=1}^H \, I_{[h \ne h']} \, \frac{N_h \, N_{h'}}{n_h \, n_{h'}} \, \alpha_h \, \alpha_{h'} \, \mu_h \, \mu_{h'} \,  \sum_{j=1}^n \sum_{j'=1}^n \, I_{[j \in {\cal C}_h, \, j' \in {\cal C}_{h'}]} \, \cdot M_{j,j'} 
\label{EVsdr} \eeq
with $M_{j,j'}$ defined on the right-hand side of (\ref{Qalt.D})
or $M_{j,j'}$ can equivalently be replaced (as in (\ref{Vsdr3})) 
by 
$$  \Big[ I_{[\gamma_j=\gamma_{j'}, \, \gamma_{j+1}=\gamma_{j'+1} ]} \, - \, I_{[\gamma_j=\gamma_{j'+1}]}\, + \, \frac{1}{2}\, I_{[\gamma_j =\gamma_{j'}, \, \gamma_{j+1} \ne \gamma_{j'+1}]} \, + \, \frac{1}{2}\, I_{[\gamma_j \ne \gamma_{j'}, \, \gamma_{j+1} = \gamma_{j' +1}]} \, \Big] \, \Big\} $$ 

\noi The true expected stratified-SRS variance to compare (\ref{EVsdr}) against is still $v_{\tt tru}$ given in (\ref{Vtru}).

\section{{\tt R} Code for $\hat{V}^{\tt SDR}$ Relative Bias and Variance} \label{rcode}

Our computations regarding SDR variance estimates 
were done with several {\tt R} functions in a package {\tt SDR} available to readers at \url{ https://github.com/talkdatatome/SDR} that are described in the following paragraphs. The vignette at the github site shows the use of these functions and some tests to show they agree when they are supposed to. In addition, the script {\tt SDRpkg.Rlog} script which is summarized in Supplement~\ref{D-EVSDR} emphasizes that the expected SDR variance from the formulas and functions is correct for $D$-cycled SDR estimates for all $n, D$, and that generally these expectations are almost exactly the same for different values of $D$, and the exact $D=1$ computation by {\tt VSDR.Gen} for all $n$ can be reasonably approximated by the $D=1$ computation in {\tt RelSDRBias} that effectively assumes $n \le m$.

Calculation of the expected SDR variance-estimator  formulas (\ref{VsdrG})--(\ref{Qsdr2}) for the total of a weighted survey attribute {\tt ywt} is given in function {\tt RelSDRBias}.  This function with default arguments calculates the expected SDR variance estimate with $D=1$, depending only on parameters $\{N_h, \, n_h, \, \mu_h, \, \sigma_h, \, \alpha_h\}_{h=1}^H$. These formulas reflect the stratified SRS sampling design and therefore require $n_h \ge 1$ for all $h=1,\ldots, H$.  To calculate the expected SDR estimates of domain variances in the more general formula (\ref{Vsdr2}) applicable for all $n, D$, there are two functions, {\tt VSDR.Gen} ands {\tt VSDRbias}. They calculate the same thing, the expected SDR Variance as given in formula (\ref{EVsdr}) for a specified $D$ and permutations $a_j^{(d)}$ or $\gamma_j$, by two slightly different combinatorial methods.  

The function {\tt PermSDR} computes $\hat{V}^{\tt SDR}$ for all $n, D$ in terms of weighted attributes {\tt ywt} total, without recourse to Hadamard matrices, based on a sample split (in order) into strata with sample-sizes given in the vector argument {\tt nhvec}, with the strata $1$ to $H$ permuted according to the permutation {\tt Str.perm}. This function, like  {\tt VSDRbias} mentioned above, uses a sparse-matrix representation {\tt Mmat} of the $n \times n$ matrix with $(j,j')$  entry given by the right-hand side of (\ref{Qsdr2})). This {\tt Mmat} matrix is calculated by a function {\tt Mcalc(n, D)}.

A function {\tt VarSDRVar} implements formula (\ref{VSDRvar}) for $var(\hat{V}^{\tt SDR})$ based on an SRS sample from an {\tt iid\/} superpopulation, as well as a function {\tt GpCalc} to identify the $G$ distinct cycling groups as in (\ref{Ygdef}).

\newpage
\section*{\Large\bf  Supplement}

\setcounter{section}{19}
\subsection{Multi-cycle Indexing in the ACS} \label{ACScycles}

The indexing (\ref{fdefD}) used for SDR in ACS has parameters $R=80, \, m=78, \, D=10$. Navarro (2002), Sukasih and Jang (2003), 
and Ash (2014) all describe in words a similar idea, based on successive cycles modified from arithmetic progressions with progressively longer spans, for constructing $10$ cycles $\mathbf{a}^{(d)}$ of row-indices from the row-numbers $\{1,\ldots, 80\}$ of an $80\times 80$ Hadamard matrix, omitting rows $1, 41$ .

 Simple {\tt R} code to build the $d$'th cycle from arithmetic progressions of span $d$, re-cycling as necessary to use all indices  in ${\cal A} = \{2,\ldots, 40, 42,\ldots, 80\}$, as given in Appendix~\ref{cycle10} is: \vspace{-2mm}
\begin{verbatim}PermArray = array(0, c(78,10))
PermArray[,1] = c(2:40,42:80)
for(j in 2:10) {
       cyc = NULL
       for(k in 1:j) cyc = c(cyc, seq(1+k,80,by=j))
       PermArray[,j] = setdiff(cyc,41) }
\end{verbatim}
The resulting columns $\mathbf{a}^{(d)}, \; d=1,\ldots, 10$ are : 

$\mathbf{a}^{(1)} = $ (2, 3, 4, 5, 6, 7, 8, 9, 10, 11, 12, 13, 14, 15, 16, 17, 18, 19, 20, 21, \\
22, 23, 24, 25, 26, 27, 28, 29, 30, 31, 32, 33, 34, 35, 36, 37, 38, 39, 40, 42, \\
43, 44, 45, 46, 47, 48, 49, 50, 51, 52, 53, 54, 55, 56, 57, 58, 59, 60, 61, 62, \\
63, 64, 65, 66, 67, 68, 69, 70, 71, 72, 73, 74, 75, 76, 77, 78, 79, 80) 

$\mathbf{a}^{(2)} = $ (2, 4, 6, 8, 10, 12, 14, 16, 18, 20, 22, 24, 26, 28, 30, 32, 34, 36, 38, \\
40, 42, 44, 46, 48, 50, 52, 54, 56, 58, 60, 62, 64, 66, 68, 70, 72, 74, 76, 78,\\
 80, 3, 5, 7, 9, 11, 13, 15, 17, 19, 21, 23, 25, 27, 29, 31, 33, 35, 37, 39,  \\
 43, 45, 47, 49, 51, 53, 55, 57, 59, 61, 63, 65, 67, 69, 71, 73, 75, 77, 79)

$\mathbf{a}^{(3)} = $ (2, 5, 8, 11, 14, 17, 20, 23, 26, 29, 32, 35, 38, 44, 47, 50, 53, 56, 59, \\ 
62, 65, 68, 71, 74, 77, 80, 3, 6, 9, 12, 15, 18, 21, 24, 27, 30, 33, 36,  39, \\ 
42, 45, 48, 51, 54, 57, 60, 63, 66, 69, 72, 75, 78, 4, 7, 10, 13, 16, 19, 22,  \\ 
25, 28, 31, 34, 37, 40, 43, 46, 49, 52, 55, 58, 61, 64, 67, 70, 73, 76, 79)

$\mathbf{a}^{(4)} = $ (2, 6, 10, 14, 18, 22, 26, 30, 34, 38, 42, 46, 50, 54, 58, 62, 66, 70, \\
74, 78, 3, 7, 11, 15, 19, 23, 27, 31, 35, 39, 43, 47, 51, 55, 59, 63, 67, 71, \\ 
75, 79, 4, 8, 12, 16, 20, 24, 28, 32, 36, 40, 44, 48, 52, 56, 60, 64, 68, 72,  \\
76, 80, 5, 9, 13, 17, 21, 25, 29, 33, 37, 45, 49, 53, 57, 61, 65, 69, 73, 77)

$\mathbf{a}^{(5)} = $ (2, 7, 12, 17, 22, 27, 32, 37,  42, 47, 52, 57, 62, 67, 72, 77, 3, 8, 13,\\ 
18, 23, 28, 33, 38, 43, 48, 53, 58, 63, 68, 73, 78, 4, 9, 14, 19, 24, 29,  34, \\
39, 44, 49, 54, 59, 64, 69, 74, 79, 5, 10, 15, 20, 25, 30, 35, 40, 45, 50, 55, \\
60, 65, 70, 75, 80, 6, 11, 16, 21, 26, 31, 36, 46, 51, 56, 61, 66, 71, 76)

$\mathbf{a}^{(6)} = $ (2, 8, 14, 20, 26, 32, 38, 44, 50, 56, 62, 68, 74, 80, 3, 9, 15, 21, 27,\\
  33, 39, 45, 51, 57, 63, 69, 75, 4, 10, 16, 22, 28, 34, 40, 46, 52, 58, 64, 70,\\ 
 76, 5, 11, 17, 23, 29, 35, 47, 53, 59, 65, 71, 77, 6, 12, 18, 24, 30, 36, 42, \\
 48, 54, 60, 66, 72, 78, 7, 13, 19, 25, 31, 37, 43, 49, 55, 61, 67, 73, 79)

$\mathbf{a}^{(7)} = $ (2, 9, 16, 23, 30, 37, 44, 51, 58, 65, 72, 79, 3, 10, 17, 24, 31, 38,\\
 45, 52, 59, 66, 73, 80, 4, 11, 18, 25, 32, 39, 46, 53, 60, 67, 74, 5, 12, 19, \\
26, 33, 40, 47, 54, 61, 68, 75, 6, 13, 20, 27, 34, 48, 55, 62, 69, 76, 7, 14,\\
 21, 28,  35, 42, 49, 56, 63, 70, 77, 8, 15, 22, 29, 36, 43, 50, 57, 64, 71, 78)

$\mathbf{a}^{(8)} = $ (2, 10, 18, 26, 34, 42, 50, 58, 66, 74, 3, 11, 19, 27, 35, 43, 51, 59,  \\
67, 75, 4, 12, 20, 28, 36, 44, 52, 60, 68, 76, 5, 13, 21, 29, 37, 45, 53, 61, 69, \\
77,  6, 14, 22, 30, 38, 46, 54, 62, 70, 78, 7, 15, 23, 31, 39, 47, 55, 63, 71,  \\
79, 8, 16, 24, 32, 40, 48, 56, 64, 72, 80, 9, 17, 25, 33, 49, 57, 65, 73)

$\mathbf{a}^{(9)} = $ (2, 11, 20, 29, 38, 47, 56, 65, 74, 3, 12, 21, 30, 39, 48, 57,  66, 75, \\
4, 13, 22, 31, 40, 49, 58, 67, 76, 5, 14, 23, 32, 50, 59, 68, 77, 6, 15, 24, 33, \\
 42, 51, 60, 69, 78, 7, 16, 25, 34, 43, 52, 61, 70, 79, 8, 17, 26, 35, 44, 53, 62, \\ 
 71, 80, 9, 18, 27, 36, 45, 54, 63, 72, 10, 19, 28, 37, 46, 55, 64, 73)

$\mathbf{a}^{(10)} = $ (2, 12, 22, 32, 42, 52, 62, 72, 3, 13, 23, 33, 43, 53, 63, 73, 4, 14, \\
24, 34, 44, 54, 64, 74, 5, 15, 25, 35, 45, 55, 65, 75, 6, 16, 26, 36, 46, 56, 66, \\
 76, 7, 17, 27, 37, 47, 57, 67, 77, 8, 18, 28, 38, 48, 58, 68, 78, 9, 19, 29, 39, \\
49, 59, 69, 79, 10,  20, 30, 40, 50, 60, 70, 80, 11, 21, 31, 51, 61, 71)

The set of 10 cycles of 78 actually used in ACS indexing for SDR is generated using the SAS code by 
K.~Albright given in Navarro (2002). Running that code and renumbering rows to run from 2 to 80 (using 
the rule  $k \mapsto 1 +( (k-1) \mod 80$) yields the following 10 cycles $\mathbf{a}^{*(d)}, \; d=1,\ldots, 10$:

$\mathbf{a}^{*(1)} = $ (2, 3, 4, 5, 6, 7, 8, 9, 10, 11, 12, 13, 14, 15, 16, 17, 18, 19, 20, 21, \\
22, 23, 24, 25, 26, 27, 28, 29, 30, 31, 32, 33, 34, 35, 36, 37, 38, 39, 40, 42, \\
43, 44, 45, 46, 47, 48, 49, 50, 51, 52, 53, 54, 55, 56, 57, 58, 59, 60, 61, 62, \\
63, 64, 65, 66, 67, 68, 69, 70, 71, 72, 73, 74, 75, 76, 77, 78, 79, 80)

$\mathbf{a}^{*(2)} = $ (3, 5, 7, 9, 11, 13, 15, 17, 19, 21, 23, 25, 27, 29, 31, 33, 35, 37, 39, 2, \\
4, 6, 8, 10, 12, 14, 16, 18, 20, 22, 24, 26, 28, 30, 32, 34, 36, 38, 40, 43, \\
45, 47, 49, 51, 53, 55, 57, 59, 61, 63, 65, 67, 69, 71, 73, 75, 77, 79, 42, 44, \\
46, 48, 50, 52, 54, 56, 58, 60, 62, 64, 66, 68, 70, 72, 74, 76, 78, 80)

$\mathbf{a}^{*(3)} = $ (3, 6, 9, 12, 15, 18, 21, 24, 27, 30, 33, 36, 39, 2, 5, 8, 11, 14, 17, 20, \\ 
23, 26, 29, 32, 35, 38, 4, 7, 10, 13, 16, 19, 22, 25, 28, 31, 34, 37, 40, 43,  \\
46, 49, 52, 55, 58, 61, 64, 67, 70, 73, 76, 79, 42, 45, 48, 51, 54, 57, 60, 63,  \\
66, 69, 72, 75, 78, 44, 47, 50, 53, 56, 59, 62, 65, 68, 71, 74, 77, 80)

$\mathbf{a}^{*(4)} = $ (5, 9, 13, 17, 21, 25, 29, 33, 37, 2, 6, 10, 14, 18, 22, 26, 30, 34, 38, 3, \\ 
7, 11, 15, 19, 23, 27, 31, 35, 39, 4, 8, 12, 16, 20, 24, 28, 32, 36, 40, 45, \\
49, 53, 57, 61, 65, 69, 73, 77, 42, 46, 50, 54, 58, 62, 66, 70, 74, 78, 43, 47, \\
51, 55, 59, 63, 67, 71, 75, 79, 44, 48, 52, 56, 60, 64, 68, 72, 76, 80)

$\mathbf{a}^{*(5)} = $ (6, 11, 16, 21, 26, 31, 36, 2, 7, 12, 17, 22, 27, 32, 37, 3, 8, 13, 18, 23, \\
28, 33, 38, 4, 9, 14, 19, 24, 29, 34, 39, 5, 10, 15, 20, 25, 30, 35, 40, 46,\\
51, 56, 61, 66, 71, 76, 42, 47, 52, 57, 62, 67, 72, 77, 43, 48, 53, 58, 63, 68, \\
73, 78, 44, 49, 54, 59, 64, 69, 74, 79, 45, 50, 55, 60, 65, 70, 75, 80)

$\mathbf{a}^{*(6)} = $ (3, 9, 15, 21, 27, 33, 39, 2, 8, 14, 20, 26, 32, 38, 7, 13, 19, 25, 31, 37, \\
6, 12, 18, 24, 30, 36, 5, 11, 17, 23, 29, 35, 4, 10, 16, 22, 28, 34, 40, 43, \\
49, 55, 61, 67, 73, 79, 42, 48, 54, 60, 66, 72, 78, 47, 53, 59, 65, 71, 77, 46, \\
52, 58, 64, 70, 76, 45, 51, 57, 63, 69, 75, 44, 50, 56, 62, 68, 74, 80)

$\mathbf{a}^{*(7)} = $ (8, 15, 22, 29, 36, 4, 11, 18, 25, 32, 39, 7, 14, 21, 28, 35, 3, 10, 17, 24, \\
31, 38, 6, 13, 20, 27, 34, 2, 9, 16, 23, 30, 37, 5, 12, 19, 26, 33, 40, 48, \\
55, 62, 69, 76, 44, 51, 58, 65, 72, 79, 47, 54, 61, 68, 75, 43, 50, 57, 64, 71, \\
78, 46, 53, 60, 67, 74, 42, 49, 56, 63, 70, 77, 45, 52, 59, 66, 73, 80)

$\mathbf{a}^{*(8)} = $ (9, 17, 25, 33, 2, 10, 18, 26, 34, 3, 11, 19, 27, 35, 4, 12, 20, 28, 36, 5, \\
13, 21, 29, 37, 6, 14, 22, 30, 38, 7, 15, 23, 31, 39, 8, 16, 24, 32, 40, 49, \\
57, 65, 73, 42, 50, 58, 66, 74, 43, 51, 59, 67, 75, 44, 52, 60, 68, 76, 45, 53, \\
61, 69, 77, 46, 54, 62, 70, 78, 47, 55, 63, 71, 79, 48, 56, 64, 72, 80)

$\mathbf{a}^{*(9)} = $ (3, 12, 21, 30, 39, 2, 11, 20, 29, 38, 10, 19, 28, 37, 9, 18, 27, 36, 8, 17, \\
26, 35, 7, 16, 25, 34, 6, 15, 24, 33, 5, 14, 23, 32, 4, 13, 22, 31, 40, 43, \\
52, 61, 70, 79, 42, 51, 60, 69, 78, 50, 59, 68, 77, 49, 58, 67, 76, 48, 57, 66, \\
75, 47, 56, 65, 74, 46, 55, 64, 73, 45, 54, 63, 72, 44, 53, 62, 71, 80)

$\mathbf{a}^{*(10)} = $ (11, 21, 31, 2, 12, 22, 32, 3, 13, 23, 33, 4, 14, 24, 34, 5, 15, 25, 35, 6, \\
16, 26, 36, 7, 17, 27, 37, 8, 18, 28, 38, 9, 19, 29, 39, 10, 20, 30, 40, 51, \\
61, 71, 42, 52, 62, 72, 43, 53, 63, 73, 44, 54, 64, 74, 45, 55, 65, 75, 46, 56, \\
66, 76, 47, 57, 67, 77, 48, 58, 68, 78, 49, 59, 69, 79, 50, 60, 70, 80

\subsection{Figures Based on $D$-cycling in Additional Superpopulations} \label{morepops}

\begin{figure}[H]
    \begin{center}
        \includegraphics[width=\textwidth]{"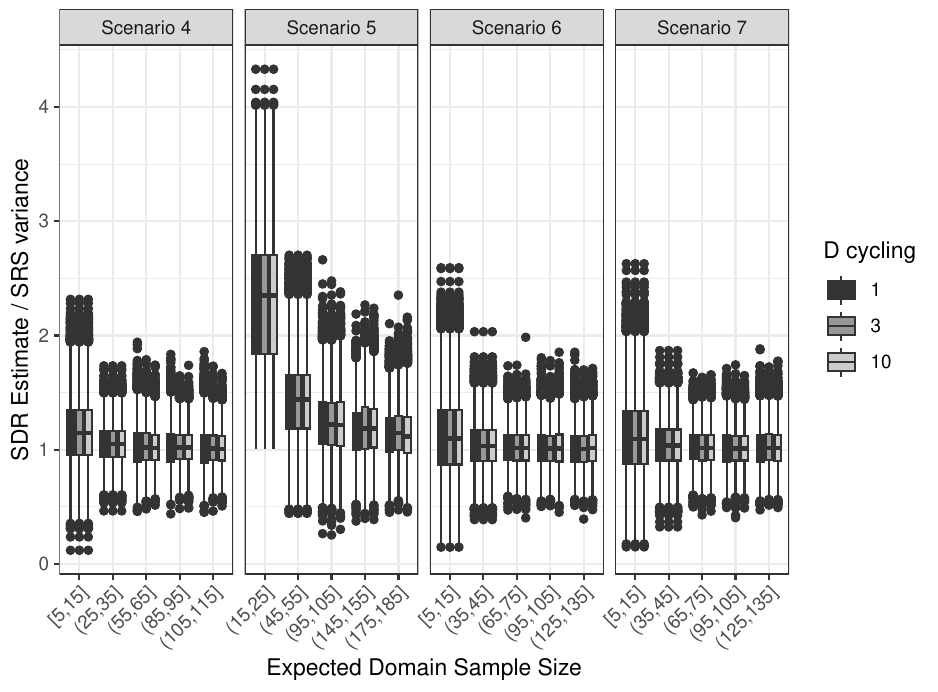"}
    \caption{Boxplots for ratios $\hat{V}^{\tt SDR}/V^{\tt tru}$ of SDR variance estimates over true stratified SRS variances for 10,000 stratified SRS samples drawn from 4 superpopulations generated randomly as in Sec.~\ref{sims}, for $D$-cycling parameters $1, 3, 10$. The superpopulations in this figure are different from those of Figure~\ref{Dcycling_plot}, but the pattern is the same: very small differences aross $D$.}
    \label{Dcycling.supplment.plot}
    \end{center}
\end{figure}

The behavior of SDR variance estimates was seen for various sample sizes to depend remarkably little on the cycling parameter $D$ used in defining SDR. We show here the similar exhibits for four additional superpopulations generated randomly as in Sec.~\ref{sims}.  We have examined  many other superpopulations and stratified-sampling designs 
with the same pattern of results, strengthening our confidence in the recommendation that the bias of SDR variance estimates depends only minimally on the choice of $D$.

\subsection{Permutation of Strata} \label{perm}

Because formula (\ref{RelBias}) suggests that alternation of $q_h = p_h \mu_h \alpha_h/n_h$ across successive strata accentuates the relative bias of SDR variance estimators, it is possible that re-ordering the strata -- while leaving the ordering of units within strata unchanged -- before implementing SDR might reduce that bias. The optimal reordering would be specific to the outcome variable $y$, and since the associated parameters $\mu_h, \, \sigma_h^2$ would generally not be known, any reordering should not involve them. One easily implemented plan is to permute the strata randomly and estimate variance by the median across permutations of the SDR variance $\hat{V}^{\tt SDR}$, and this subsection investigates the effects of such an estimation strategy. 

\begin{figure}[H]
    \begin{center}
        \includegraphics[width=\textwidth]{"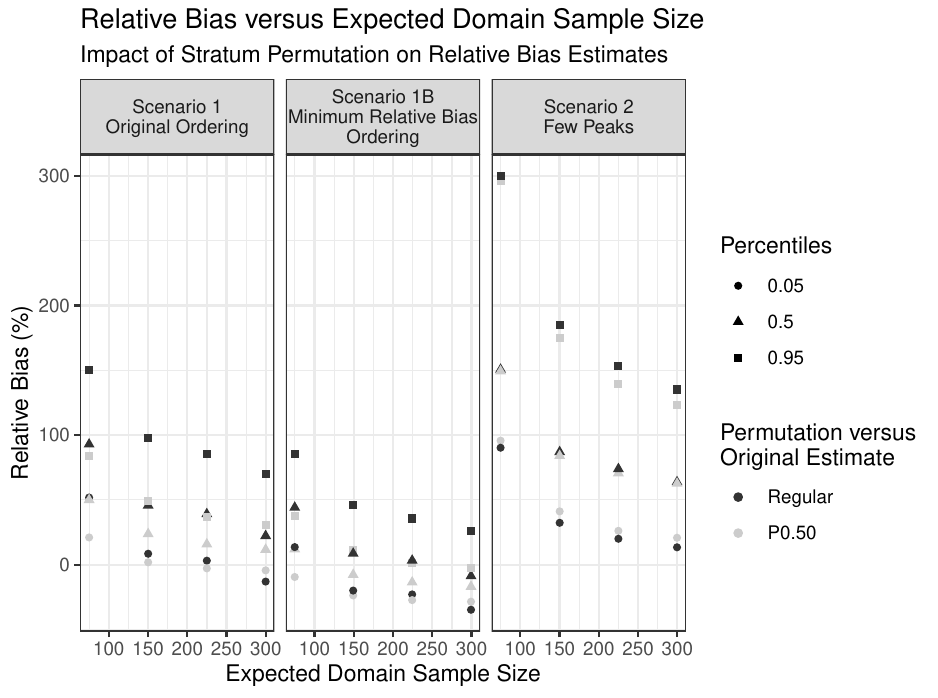"}
    \caption{Quantiles ($0.05, 0.5$, and $0.95$, respectively denoted by circle, triangle and square) of relative bias for $\hat{V}^{\tt SDR}$ estimates versus medians (black versus gray points) of estimates over 1000 random permutations of stratum numbers. Quantiles are computed across 10,000 samples from each of 3 superpopulations (in the three panel) at 5 different domain sample sizes. Superpopulation Scenarios are successively characterized by dramatic alternation in $q_h$, by essentially no alternation, and by a pattern of $q_h$ with isolated peaks. Variability of estimates is smaller in all three panels for the median-over-permutations estimates, but the relative biases themselves are noticeably smaller only in Scenario 1.}
    \label{perm.plot}
    \end{center}
\end{figure}

Consider the random-permutation strategy for superpopulations with 30 strata, in which $10^4$ stratified SRS samples are drawn for 3 superpopulations: Scenario 1 (`Alternating') with  $q_h$ purely alternating between values $0.0006, 0.024$, Scenario 1B (with strata ordered so that those same $q_h$ values are increasing) and Scenario 2 (`Isolated Peaks') in which the $q_h$ have 3 isolated peaks (of sizes roughly $0.102, 0.030, 0.009$ with all other $q_h$ values very small $< 0.002$). Parameters in these scenarios are constructed to give the same expected domain sample sizes for each $n$. Figure~\ref{perm.plot} displays the 5th, 50th and 95 percentiles across samples of the relative bias of the original $\hat{V}^{\tt SDR}$ and of the variance estimate obtained as the median of $\hat{V}^{\tt SDR}$ values for 1000 random permutations.

Over the range of sample sizes in Figure~\ref{perm.plot}, and in all three superpopulations, SDR variance estimates are less variable when the single $\hat{V}^{\tt SDR}$ estimate is replaced by the median over random permutations of strata. the variability in the relative bias estimates drops when one takes the median of a random set of stratum level permutations. In Scenario 1 in the original ordering, the relative bias of the median-over-permutations estimate is also smaller relative to the original estimates. However, the monotone ordering of Scenario 1B makes the relative bias (\ref{RelBias}) particularly small, so it is not surprising that taking the median-across-permutations actually raises the relative bias. The ordering of strata in Scenario 2 matters much less, since only very few permutations place the few peaks in $q_h$ next to one another. 

\subsection{Expected SDR Variance  with Different $D$ for StratSRS Designs} \label{D-EVSDR}

The interpretable formula (\ref{VSDR2}) for expected SDR variance applied precisely only to stratified SRS designs with $n \le m$. It was used in Section~\ref{SDR.Bias} to characterize typical SDR biases from small and moderate-sized StratSRS surveys drawn from {\tt iid\/} superpopulations with cycling parameter $D=1$, and the $D=1$ case is the one that has been most studied empirically in the SDR literature. Yet the Census Bureau, which is undoubtedly the mostly consistent user of SDR variance estimation, implements SDR with $D=10$ in its large surveys (ACS and CPS). In  those surveys, and in other real applications of the SDR method, the sample size $n = \sum_{h=1}^H \, n_h$ is typically large, even when restricted to states or large counties, but the expected sample size $E(n_A) \, = \,\sum_{h=1}^H \, \alpha_h \, n_h$ in domains $A$ of interest will often not be large. 

Now we have a method, described in Appendix~\ref{D.relBias} and implemented in the {\tt R} function {\tt VSDR.Gen}, of exactly calculating SDR bias for domain total estimates based on StratSRS designs (from {\tt iid\/} superpopulations) when calculated with arbitrary $n$ and any $D$. So it is interesting to compare these biases, or equivalently the expected SDR variances $E(\hat{V}^{\tt SDR})$, across $D$. In this supplementary subsection, we illustrate some of the many calculations we have done with the function {\tt VSDR.Gen} to compare the expected SDR variances across $D=1,\ldots, 10$.

The brief summary of what we have found in our calculations with different $D$ on the same parameter sets ${\bf P} = (N_h, n_h, \mu_h, \sigma_h, \alpha_h, \;~ h=1,\ldots, H)$ is as follows:  
\par (i) the expected variances are almost always close, in the sense that the relative biases are within a few percent except when $E(n_A)$ is large and the variances themselves are small (so that the CV's of estimates are less than 10\% or so); 
\par (ii) the close correspondence between the $E(\hat{V}^{\tt SDR})$ values calculated (correctly) by {\tt VSDR.Gen} for different $D$ extends also to the simple formula~\ref{VSDR2} calculated as though $n \le m$ and implemented in the {\tt R} function {\tt RelSDRBias}.

These general conclusions are illustrated with computed results as usage examples in the {\tt SDRpkg.Rlog} script and in the vignette supplied for the {\tt R} package {\tt SDR} at  \url{ https://github.com/talkdatatome/SDR}.

\end{document}